\documentclass[11pt, reqno]{amsart}
\usepackage{eurosym}
\usepackage{amsfonts}
\usepackage{layout}
\usepackage[top=2cm, bottom=2cm, left=2.5cm, right=2.5cm]{geometry}
\usepackage{amssymb}
\usepackage{amsmath, amsopn}
\usepackage{pmboxdraw}
\usepackage{amsthm}
\usepackage[colorlinks,linkcolor=blue, citecolor=blue]{hyperref}
\usepackage[noadjust]{cite}

\makeatletter

\@addtoreset{equation}{section}
\makeatother

\newcommand{\QATOP}[2]{\genfrac{}{}{0pt}{}{#1}{#2}}

\begin{document}
\title{Husimi Q-functions attached to hyperbolic Landau levels}
\author{$^{\star}$Z. Mouayn, $^{\flat}$H. Chhaiba, $^\ddagger$H. Kassogue, $^{\Upsilon}$P. K. Kikodio}


\maketitle

\begin{center}
	\begin{scriptsize}
		\textit{$^{\star}$ Department of Mathematics, Faculty of Sciences and Technics (M’Ghila),\\
			Sultan Moulay Slimane University, P.O. Box. 523, B\'eni Mellal, Morocco.
			mouayn@usms.ma\\
			$^{\flat}$ Residence num 408 Hassan 2 avenue appt 6, Diour Jamaa, Rabat, Morocco.
			chhaiba.hassan@gmail.com\\
			$^\ddagger$ High Institute of Applied Technologies (TechnoLAB-ISTA), Bamako, Mali.
			hamidoukass@gmail.com\\
			$^{\Upsilon}$ 13, Market avenue, Mongt-Ngafula, Kinshasa, DR Congo.
			kayupepatrick@gmail.com}
	\end{scriptsize}
\end{center}

\begin{abstract}
We are concerned with a phase-space probability distribution which is known
as Husimi $Q$-function of a density operator with respect to a set of
coherent states $\vert\widetilde{\kappa}_{z,B,R,m}\rangle$ attached to an $m$%
th hyperbolic Landau level and labeled by points $z$ of an open disk of
radius $R$, where $B>0$ is proportional to a magnetic field strength. For a
density operator representing a projector on a Fock state $\left\vert
j\right\rangle$ we obtain the $Q_{j}$ distribution and discuss some of its
basic properties such as its characteristic function and its main
statistical parameters. We achieve the same program for the thermal density
operator (mixed states) of the isotonic oscillator for which we establish a
lower bound for the associated thermodynamical potential. We recover most of
the results of the Euclidean setting (flat case) as the parameter $R$ goes
to infinity by making appeal to asymptotic formulas involving orthogonal
polynomials and special functions. As a tool, we establish a summation
formula for the special Kamp\'{e} de F\'{e}riet function $\digamma
_{2:0:0}^{1:2:2}$.
\end{abstract}
\section{Introduction}

Coherent states (CS) are an overcomplete family of normalized \textit{ket}
vectors $\left\vert \zeta \right\rangle $, $\zeta \in X$, in a Hilbert space 
$\mathcal{H}$ corresponding to a specific quantum model and provide $%
\mathcal{H}$ with the following resolution of the identity 
\begin{equation}
1_{\mathcal{H}}=\int\limits_{X}\left\vert \zeta \right\rangle \left\langle
\zeta \right\vert d\mu \left( \zeta \right).
\end{equation}%
Here, $X$ represents the phase-space domain and $d\mu \left( \zeta \right) $
the associated integration measure on $X$. These CS have long been known for
the harmonic oscillator, whose properties have been used as a model for more
quantum systems \cite{Gazeau}. In a such system of states the density
operator $\hat{\rho}$ for an arbitrary (pure or mixed) state can be
represented by the classical Husimi $Q$-function \cite{Husimi}: 
\begin{equation}
Q_{\hat{\rho}}(\zeta ):=\left\langle \zeta \right\vert \hat{\rho}\left\vert
\zeta \right\rangle.  \label{1.2}
\end{equation}%
In particular, for a pure state of the form $\hat{\rho}=\left\vert \phi
\right\rangle \left\langle \phi \right\vert $ where $\left\vert \phi
\right\rangle \in \mathcal{H}$ is an arbitrary vector, the $Q$-function can
also be expressed as 
\begin{equation}
Q_{\left\vert \phi \right\rangle \left\langle \phi \right\vert }(\zeta
)=\left( \mathcal{N}_{\zeta }\right) ^{-1}\left\vert W\left[ \phi \right]
\left( \zeta \right) \right\vert ^{2}
\end{equation}%
in terms of the associated \textit{coherent state transform W }defined from $%
\mathcal{H}$ into a specific $L^{2}$ subspace of $X.$ Here, $\mathcal{N}%
_{\zeta }$ is a factor ensuring the normalization condition $\langle \zeta
\left\vert \zeta \right\rangle =1.$ For a general density operator $\hat{\rho%
},$ the function $Q_{\hat{\rho}}(\zeta )$ is normalized, since 
\begin{equation}
1=\mathrm{Tr}\left[ \hat{\rho}\right] =\int\limits_{X}Q_{\hat{\rho}}(\zeta )d\mu
(\zeta ).
\end{equation}%
As $Q_{\hat{\rho}}$ is introduced in a way that guarantees it to be non
negative, then it allows the representation of quantum states by a \textit{%
probability distribution} in the phase space $X$. It provides, equivalently
to the Glauber-Surdashan or Wigner representations, a basis for a formal
equivalence between the quantum and classical descriptions of optical
coherence \cite{Glauber}.


In this paper, we are concerned with the Husimi function of a density
operator with respect to a set of generalized CS (GCS) $\left\vert 
\widetilde{\kappa }_{z,B,R,m}\right\rangle $ belonging to the Hilbert space $%
L^{2}\left( \mathbb{R}_{+}\right) ,$ labeled by points $z\in \mathbb{D}_{R}
=\left\{ \zeta \in \mathbb{C},\left\vert \zeta \right\vert <R\right\} $, $R>0
$ and attached to a higher $m$th hyperbolic Landau level in the presence of
a magnetic field whose strength is proportional to $B>0$. We first introduce
GCS via a group theoretical method by displacing an $m$th Laguerre function
by mean of a square integrable irreducible unitary representation of the
affine group. These GCS are then used to obtain the Husimi distribution, $%
Q_{j}^{(B,R,m)}$, for a projector operator $\hat{\rho}_{j}=\left|j\right%
\rangle\left\langle j \right|$ on a Fock state $\left\vert j\right\rangle $,
which is known as a pure state. For this distribution we write down the
characteristic function from which we derive the main value and the
variance. As in the Euclidean setting \cite{Mouayn7} this distribution may
be useful for tackling the hyperbolic version of Ginibre-type processes \cite%
{Shirai}. Next, from $Q_{j}^{(B,R,m)}$ we deduce the Husimi function $%
Q_{\beta}^{(B,R,m)}$ for the thermal density operator (heat semi-group) $%
\hat{\rho}_{\beta}$ of the Hamiltonian of the isotonic oscillator. Since $%
Q_{j}^{(B,R,m)}$ may be also viewed as a lower symbol for $\hat{\rho}_{\beta}
$ we apply a Berezin-Lieb inequality \cite{Berezin} (with a specific choice
of a convex function) to obtain a lower bound for the thermodynamical
potential associated with this Hamiltonian. We also establish a formula for
the characteristic function of $Q_{\beta}^{(B,R,m)}$, from which we derive
the mean and the variance parameters. Finally, we recover most of the
results \cite{Mouayn7} of the flat case as the radius parameter $R$ goes to
infinity. As a tool we establish a summation formula for the special Kamp\'e de F\'eriet function $F^{1:2:2}_{2:0:0}$.

The paper is organized as follows. In Section \ref{section2}, we recall some
notations and facts about orthogonal polynomials and special functions we
will be using. In section \ref{section3}, we discuss the construction of GCS
labeled by elements of the affine group as well as their relationship to
hyperbolic Landau levels. Section \ref{section4} deals with the
Q-representation for a pure state as well as the corresponding
characteristic function and the main statistical parameters. The same task
is acheived in Section \ref{section5} for the the thermal density operator
of the isotonic oscillator. For the latter one, we also give a lower bound
for the thermodynamical potential. In Section \ref{section6}, we establish a
summation formula for the Kamp\'e de F\'eriet function $\digamma
_{2:0:0}^{1:2:2}$. Section \ref{section7} is devoted to some concluding
remarks. Proofs of our results are detailed in appendices.

\section{Some notations and definitions}
\label{section2} This section collects the basic notations and definitions
of special functions and orthogonal polynomials used in the rest of the
paper. The reader may proceed to Section \ref{section3} and refer back here
as necessary. For more details on the theory of these functions we refer to 
\cite{Chihara,Szego,Erdelyi,Srivastava3,Appel}. \newline
\newline
\textbf{1.} For $a\in\mathbb{C}$, the shifted factorial or Pochhammer symbol
is defined by 
\begin{equation}  \label{pochhamer}
(a)_k = a(a+1)\dots(a+k-1),\qquad k\in\mathbb{N}
\end{equation}
where by convention $(a)_0=1$. When $a=-n$ with $n\in\mathbb{N}^*=\mathbb{N}%
-\{0\}$, 
\begin{equation}
(-n)_k = \left\{ 
\begin{array}{cl}
\frac{(-1)^kn!}{(n-k)!},\qquad & 0\leq k\leq n \\ 
0, \qquad& k>n%
\end{array}
\right. .  \label{Pochhammer-n}
\end{equation}
\textbf{2.} For $a\in\mathbb{C}$ and $k \in\mathbb{N}$, the binomial
coefficient is defined by 
\begin{equation}
\binom{a}{k} =\frac{a(a-1)\dots(a-k+1)}{k!} = \frac{(-1)^k(-a)_k}{k!}
\end{equation}
\newline
\textbf{3.} For $z\in\mathbb{C}$, the gamma function is defined by 
\begin{equation}  \label{gamma}
\Gamma(z) = \int\limits_{0}^{\infty} t^{z-1} e^{-t} dt,\qquad {\rm Re}%
\,z>0.
\end{equation}
Note that $\Gamma(n+1)=n!$ if $n\in\mathbb{N}$, and 
\begin{equation}
(a)_k = \frac{\Gamma(a+k)}{\Gamma(a)}  \label{PochhammerGG}
\end{equation}
if $a\in\mathbb{C}\setminus \mathbb{Z}_-$. \newline
\textbf{4.} For $a,b \in\mathbb{C}$ such that $\mbox{Re}\,a,\mbox{Re}\,b>0$,
the beta function is defined by 
\begin{equation}
\mathcal{B}(a,b)=\int\limits_{0}^{1} t^{a-1} (1-t)^{b-1} dt =\frac{%
\Gamma(a)\Gamma(b)} {\Gamma (a+b)}.  \label{Beta}
\end{equation}
\textbf{5.} For $a_1,\dots,a_p\in\mathbb{C}$ and $c_1,\dots,c_q\in \mathbb{C}%
\setminus \mathbb{Z}_-$, the generalized hypergeometric function is defined
by the series 
\begin{equation}
{}_p\digamma_q\left( {\QATOP{a_1,...,a_p}{c_1,...,c_q}}\big|z\right) =
\sum_{k=0}^{+\infty} \frac{(a_1)_k\cdots(a_p)_k}{(c_1)_k\cdots(c_q)_k} \frac{%
z^{k}}{k!},  \label{hypergeometric}
\end{equation}
which terminates whenever at least one of the $p$ parameters $a_i$ equals $%
-1,-2,-3,\dots$. It converges for $|z|<\infty$ if $p\leq q$ or for $|z|<1$
if $p=q+1$ and it diverges for all $z\neq0$ if $p>q+1$. Special cases of
this function are the Gauss hypergeometric function ${}_2\digamma_1({\QATOP{%
a,b}{c}}|z)$, the confluent hypergeometric function ${}_1\digamma_1({\QATOP{a%
}{c}}|z)$ and the binomial series ${}_1\digamma_0({\QATOP{a}{-}}|z)$. The
later one reduces to $(1-z)^{-a}$ if $|z|<1$.\newline
\textbf{6.} For $a\in\mathbb{R}$, the series 
\begin{equation}  \label{IBessel}
I_{a}(z) = \sum_{k=0}^{+\infty} \frac{\left(\frac{1}{2}z\right)^{a+2k}} {%
k!\Gamma(a+k+1)}\text{, } \qquad z\in \mathbb{C}
\end{equation}
is called the modified Bessel function of the first kind and order $a$. 
\newline
\textbf{7.} The Jacobi polynomial of parameters $a$ and $b$ is defined by 
\begin{equation}
P_{n}^{(a,b)}(x)=2^{-n}\sum_{k=0}^{n}\binom{n+a}{k}\binom{n+b}{n-k}%
(x+1)^{k}(x-1)^{n-k},\qquad a,b>-1  \label{jacobi}
\end{equation}
and can be expressed in terms of the terminating ${}_2\digamma_1$-sum as 
\begin{equation}
P_{n}^{(a,b)}(x)= \frac{\Gamma (b+n+1)}{n!\Gamma(b+1)}\left(\frac{x-1}{2}%
\right)^{n} {}_{2}\digamma_{1}\left({\QATOP{-n,-(a+n)}{b+1}} \bigg| \frac{x+1%
}{x-1}\right).  \label{2F1andJacobi}
\end{equation}
\textbf{8.} The Laguerre polynomial of parameter $a$ is defined by 
\begin{equation}  \label{laguerre}
L_{n}^{\left( a \right) }\left(x\right) =\sum_{k=0}^{n}(-1)^{k}\binom{n+a}{%
n-k}\frac{x^{k}}{k!},\qquad a >-1
\end{equation}
and can be expressed in terms of the terminating ${}_1\digamma_1$-sum as 
\begin{equation}
_{1}\digamma_{1}\left( {\QATOP{-n}{a +1}}\Big|x\right) =\frac{n!}{(a +1)_{n}}%
L_{n}^{(a )}(x).  \label{1F1andLaguerre}
\end{equation}
\textbf{9.} For $a_1,\dots,a_p, b_1,\dots,b_q, c_1,\dots,c_k\in\mathbb{C}$
and $\alpha_1,\dots,\alpha_l, \beta_1,\dots,\beta_m,
\gamma_{1},\dots,\gamma_{n}\in \mathbb{C}\setminus \mathbb{Z}_-$, the Kamp%
\'{e} de F\'{e}riet function in two variables is defined by the series 
\begin{equation}  \label{kampe}
\digamma _{l:m:n}^{p:q:k}\left( {\QATOP{
a_{1}...a_{p};b_{1}...b_{q};c_{1}...c_{k}}{\alpha _{1}...\alpha _{l};\beta
_{1}...\beta _{m};\gamma _{1}...\gamma _{n}}}\bigg|x,y\right)
=\sum_{r,s=0}^{+\infty }\frac{\displaystyle\prod_{i=1}^{p}(a_{i})_{r+s}
\prod_{i=1}^{q}(b_{i})_{r}\prod_{i=1}^{k}(c_{i})_{s}}{\displaystyle %
\prod_{i=1}^{l}(\alpha _{i})_{r+s}\prod_{i=1}^{m}(\beta
_{i})_{r}\prod_{i=1}^{n}(\gamma _{i})_{s}}\frac{x^{r}}{r!}\frac{y^{s}}{s!}
\end{equation}
where for convergence, $p+q<l+m+1, \quad p+k<l+n+1, \quad|x|<\infty$ and $%
|y|<\infty$. A special case is the Humbert series 
\begin{equation}
\Phi_1\left({\QATOP{a,b }{c}} \Big| w, z\right) =
\digamma_{1:0:0}^{1:1:0}\left( {\QATOP{a;b;-}{c;-;-}}\bigg|w,z\right) =
\sum_{k,l=0}^{+\infty} \frac{(a)_{k+l}(b)_k}{(c)_{k+l}} \frac{w^kz^l}{k!l!};
 \qquad|w|<1, \, |z|<\infty  \label{Phi1Def}
\end{equation}
wich reduces to 
\begin{equation}
\Phi_1\left({\QATOP{a,b }{c}} \Big| w, 0\right) = {}_2\digamma_1\left({%
\QATOP{a,b }{c}} \Big| w\right)  \label{Phiw0}
\end{equation}
for $z=0$.

\section{Generalized Coherent states (GCS)}

\label{section3}

\subsection{GCS labeled by elements of the affine group}

We recall that the affine group is the set $\mathbf{G}=\mathbb{R}\times 
\mathbb{R}^{+}$, endowed with group law $\left( x,y\right) .\left( x^{\prime
},y^{\prime }\right) =\left( x+yx^{\prime },yy^{\prime }\right) $. $\mathbf{G%
}$ is a locally compact group with the left Haar measure $d\mu \left(
x,y\right) =y^{-2}dxdy$. We shall consider one of the two inequivalent
infinite dimensional irreducible unitary representations of the affine group 
$\mathbf{G}$, denoted $\pi _{+}$, realized on the Hilbert space $\mathcal{H}%
:=$ $L^{2}\left( \mathbb{R}^{+},\xi ^{-1}d\xi \right) $ as 
\begin{equation}
\pi _{+}\left( x,y\right) \left[ \varphi \right] \left( \xi \right) :=e^{%
\frac{i}{2}x\xi }\varphi \left( y\xi \right) ,\qquad \varphi \in \mathcal{%
\ H},\quad\xi \in \mathbb{R}^{+}\text{.}  \label{2.1}
\end{equation}%
This representation is square integrable since it is easy to find a vector $%
\phi _{0}\in \mathcal{H}$ such that the function $\left( x,y\right) \mapsto
\left\langle \pi _{+}\left( x,y\right) \left[ \phi _{0}\right] ,\phi
_{0}\right\rangle _{\mathcal{H}}$ belongs to $L^{2}\left( \mathbf{G},d\mu
\right) $. This condition can also be expressed by saying that the
self-adjoint operator $\delta :\mathcal{H\rightarrow H}$ defined as $\delta %
\left[ \varphi \right] (\xi )=\xi ^{-\frac{1}{2}}\varphi \left( \xi \right) $
gives 
\begin{equation}
\int\limits_{\mathbf{G}}\left\langle \varphi _{1},\pi _{+}\left( x,y\right) %
\left[ \psi _{1}\right] \right\rangle \left\langle \pi _{+}\left( x,y\right) %
\left[ \varphi _{2}\right] ,\psi _{2}\right\rangle d\mu \left( x,y\right)
=\left\langle \varphi _{1},\varphi _{2}\right\rangle \left\langle \delta ^{%
\frac{1}{2}}\left[ \varphi _{1}\right] ,\delta ^{\frac{1}{2}}\left[ \varphi
_{2}\right] \right\rangle  \label{2.2}
\end{equation}%
for all $\psi _{1},\psi _{2},\varphi _{1},\varphi _{2}\in \mathcal{H}$. The
operator $\delta $ is unbounded because $\mathbf{G}$ is not unimodular \cite%
{DM}.

Now, as in \cite{Mouayn4}, for $B>0$ and $m=0,1,\dots ,\lfloor B-1/2\rfloor $
where $\lfloor a\rfloor $ denotes the greatest integer not exceeding $a$, we
consider a set of CS labeled by elements $(x,y)\in \mathbf{G}$, which were
obtained by acting, via the representation operator $\pi _{+}\left(
x,y\right) $, on the admissible vector 
\begin{equation}
\phi _{B,m}\left( \xi \right) :=\left( \frac{\Gamma \left( 2B-m\right) }{m!}%
\right) ^{-\frac{1}{2}}\xi ^{B-m}e^{-\frac{1}{2}\xi }L_{m}^{\left(
2(B-m)-1\right) }\left( \xi \right)   \label{2.3}
\end{equation}%
where $L_{n}^{\alpha }(.)$ denotes the Laguerre polynomial defined in Eq.~\eqref{laguerre}. 
Precisely, 
\begin{equation}
\left\vert \tau _{(x,y),B,m}\right\rangle :=\pi _{+}\left( x,y\right) \left[
\phi _{B,m}\right]   \label{2.5}
\end{equation}%
and satisfy the resolution of the identity operator 
\begin{equation}
\mathbf{1}_{\mathcal{H}}=c_{B,m}\int\limits_{\mathbf{G}}d\mu \left(
x,y\right) \left\vert \tau _{(x,y),B,m}\right\rangle \left\langle \tau
_{(x,y),B,m}\right\vert   \label{2.6}
\end{equation}%
where $c_{B,m}:=2\left( B-m\right) -1$ and the Dirac's bra-ket notation $%
|\Phi \rangle \langle \Phi |$ means the rank-one operator $\phi \longmapsto
\langle \Phi ,\phi \rangle _{\mathcal{H}}.\Phi $ with $\Phi ,\phi \in 
\mathcal{H}$. In the $\xi $-coordinate, wavefunctions of the states in Eq.~%
\eqref{2.5} read 
\begin{equation}
\left\langle \xi \right\vert \tau _{(x,y),B,m}\rangle =\left( \frac{\Gamma
\left( 2B-m\right) }{m!}\right) ^{-\frac{1}{2}}\left( \xi y\right) ^{B-m}e^{-%
\frac{1}{2}\xi \left( y-ix\right) }L_{m}^{\left( c_{B,m}\right) }\left( \xi
y\right) \text{, }\qquad \xi >0.  \label{2.7}
\end{equation}%
For $m=0$, the states $\tau _{(x,y),B,0}$ coincide with the well known
affine coherent states \cite{Aslaken}. These affine states are closely
related to analytic wavelets \cite{AnalyticWavelet}.\ 

To describe the connection of these GCS with hyperbolic Landau levels we may
first identify the affine group with the Poincar\'{e} upper half-plane $%
\mathbb{H}^{2}\equiv $ $\mathbf{G}$ . Then, with vectors in Eq.~\eqref{2.5}
one can associate, as usual (\cite[p.188]{Bargmann}), the CS transform $%
\mathcal{B}_{m}:\mathcal{H}\rightarrow L^{2}\left( \mathbb{H}^{2},d\mu
_{B}\right) $ defined by \cite{Mouayn4}: 
\begin{equation}
\mathcal{B}_{m}[\phi ]\left( x,y\right) =\left( c_{B,m}\right) ^{\frac{1}{2}%
}\int\limits_{0}^{+\infty }\overline{\left\langle \xi \right\vert \tau
_{(x,y),B,m}\rangle }\phi (\xi )\xi ^{-1}d\xi   \label{2.8}
\end{equation}%
whose range is the eigenspace of the Schr\"{o}dinger operator (in suitable
units and up to an additive constant) describing the dynamics of a charged
particle moving on $\mathbb{H}^{2}$ under the action of a magnetic field of
strength proportional to $B$ 
\begin{equation}
\Delta _{B}=y^{2}\left( \partial _{x}^{2}+\partial _{y}^{2}\right)
-2iBy\partial _{x},  \label{2.9}
\end{equation}%
associated with the eigenvalue $\epsilon _{m}^{B}=(B-m)\left( 1-B+m\right) , $\quad
$m=0,1,...,\left\lfloor B-\frac{1}{2}\right\rfloor $. That is, $\mathcal{B}%
_{m}[\mathcal{H}]\equiv \left\{ \Phi \in L^{2}\left( \mathbb{H}^{2},d\mu
\right) ,\quad \Delta _{B}f=\epsilon _{m}^{B}f\right\} $. The operator $\Delta _{B}
$ is also known as $B$-weight Maass operator \cite%
{Ikeda,Oshima,Elstrodt,Patterson} which is an elliptic densely defined
operator on the Hilbert space $L^{2}(\mathbb{H}^{2},d\mu )$, with a unique
self-adjoint realization also denoted by $\Delta _{B}$. Its spectrum
consists of two parts:\textit{\ }a continuous part $\left[ 1/4,+\infty %
\right[ $, corresponding to \textit{scattering states} and the finite number
of eigenvalues $\epsilon _{m}^{B}$ each one with infinite degeneracy\ called 
\textit{hyperbolic Landau levels} \cite{Mouayn4}. Finally, the reproducing
kernel of the Hilbert space $\mathcal{B}_{m}[\mathcal{H}]$ can be obtained
from the overlapping function $\langle \tau _{w,B,m},\tau _{\zeta
,B,m}\rangle _{\mathcal{H}}$ between two CS as 
\begin{equation}
K_{m}^{B}\left( w,\zeta \right) =\alpha _{B,m}\left( \frac{\left\vert w-\bar{%
\zeta}\right\vert ^{2}}{4\Im w\Im \zeta }\right) ^{-B+m}\left( \frac{\zeta -%
\bar{w}}{w-\bar{\zeta}}\right) ^{B}{}_{2}\digamma _{1}\left( {\QATOP{-m,-m-2B%
}{2(B-m)}}\bigg|\frac{4\Im w\Im \zeta }{\left\vert w-\bar{\zeta}\right\vert
^{2}}\right) ,  \label{2.10}
\end{equation}%
$w,\zeta \in \mathbb{H}^{2}$ where $\alpha _{B,m}:=\frac{\left( -1\right)
^{m}\Gamma \left( 2B-m\right) }{m!\Gamma \left( 2B-2m\right) }$ and $%
_{2}\digamma _{1}$ being the Gauss hypergeometric function.

The finite number of the levels of infinite degeneracy increases with the
strength of the magnetic field, a property. In \cite{AoP} the first bounds
on the size of the fundamental region, which correspond to a `cell' (Pauli
exclusion principle allows only one electron per cell), have been obtained.
Note that \cite{AS} improves this estimate, so that, given a bounded region $%
\Delta \subset \mathbb{H}^{2}$ in a hyperbolic Landau level, assuming the
Pauli exclusion principle, it is possible to use these discrete states to
count the number of particles distributed on such $\Delta $.

\subsection{GCS labeled by points of the disk}

We can also write a version of the GCS in Eq.~\eqref{2.5} as states labeled
by points $z$ of the unit disk $\mathbb{D}=\left\{ z\in \mathbb{C},\text{ }%
\left\vert z\right\vert <1\right\} $ by using the inverse Cayley transform $%
\mathcal{C}^{-1}:\mathbb{D}\rightarrow \mathbf{G}$ given by 
\begin{equation}
\mathcal{C}^{-1}\left( z\right) =\left( -2\frac{\Im z}{\left\vert
1-z\right\vert ^{2}},\frac{1-\left\vert z\right\vert ^{2}}{\left\vert
1-z\right\vert ^{2}}\right) .  \label{2.12}
\end{equation}%
We precisely define GCS in $L^{2}(\mathbb{R}_{+},\xi ^{-1}d\xi )$ by setting 
\begin{equation}
\kappa _{z,B,m}:=\left( \frac{1-\bar{z}}{1-z}\right) ^{B}\pi _{+}\left( 
\mathcal{C}^{-1}\left( z\right) \right) \left[ \phi _{B,m}\right] .
\label{2.13}
\end{equation}%
Direct calculations lead to their wave functions in $\xi $-coordinates as 
\begin{equation}
\left\langle \xi \right\vert \kappa _{z,B,m}\rangle =\sqrt{\frac{m!}{\Gamma
\left( 2B-m\right) }}\frac{\left\vert 1-z\right\vert ^{2m}}{\left(
1-z\right) ^{2B}}\left( \left( 1-z\bar{z}\right) \xi \right) ^{B-m}\exp
\left( -\frac{\xi }{2}\frac{1+z}{1-z}\right) L_{m}^{\left( c_{B,m}\right)
}\left( \xi \frac{1-z\bar{z}}{\left\vert 1-z\right\vert ^{2}}\right).
\label{2.14}
\end{equation}%
By Eq.~\eqref{2.14} we recover the CS in (\cite[Eq.~(4.4)]{Mouayn4Bis}) (up to 
$\left( -1\right)^{m}$) where $\nu $ should be taken as our $B$ and we may
view Eq.~\eqref{2.13} as the analog of \textit{\textquotedblleft displacing
the vacuum}\textquotedblright\ $\left( D(z) \left\vert 0\right\rangle \right)
$ way in the construction of canonical CS for the harmonic oscillator. When $%
m=0,$ the wavefunction of GCS in Eq.~\eqref{2.14} reduces to 
\begin{equation}
\left\langle \xi \right\vert \kappa _{z,B,0}\rangle =\frac{1}{\sqrt{\Gamma
\left( 2B\right) }}\left( \frac{\left( 1-z\bar{z}\right) \xi }{\left(
1-z\right) ^{2}}\right) ^{B}\exp \left( -\frac{\xi }{2}\frac{1+z}{1-z}%
\right) ,\qquad\xi >0  \label{2.15}
\end{equation}%
and coincides with those constructed by Molanar \textit{et al} (\cite[%
Eq.(2.2)]{BM}) for the Morse potential by an algebraic way based on
supersymetry and shape invariance properties where the shape parameter may
be taken as our $B>0$. The GCS in Eq.~\eqref{2.15} where first introduced by
Nieto \textit{et al} \cite{NS} as generalized minimal uncertainty states for
the Hamiltonian of Morse potential \cite{Morse}.

Moreover, the GCS in Eq.~\eqref{2.14} may slightly be modified in order to
perform them as vectors of $L^{2}(\mathbb{R}_{+},d\xi )$ labeled by points
of the disk $\mathbb{D}_{R}:=\left\{ z\in \mathbb{C},\text{ }\left\vert
z\right\vert <R\right\} $ as 
\begin{equation}
\left\langle \xi \right\vert \widetilde{\kappa }_{z,B,R,m}\rangle :=\sqrt{%
\frac{2}{\xi}} \, \langle \xi^{2}\vert\kappa _{\frac{z}{R},BR^{2},m}\rangle
\end{equation}%
for every fixed $m=0,1,2,\dots ,\lfloor BR^{2}-\frac{1}{2}\rfloor $ provided
that $2BR^{2}>1$. These new states obey the normalization condition $\langle 
\widetilde{\kappa }_{z,B,R,m},\widetilde{\kappa }_{z,B,R,m}\rangle _{L^{2}(%
\mathbb{R}_{+},d\xi )}=1 $ and satisfy the resolution of the identity
operator as 
\begin{equation}
1_{L^{2}(\mathbb{R}_{+},d\xi )}=\int\limits_{\mathbb{D}_{R}}\left\vert 
\widetilde{\kappa }_{z,B,R,m}\right\rangle \left\langle \widetilde{\kappa }%
_{z,B,R,m}\right\vert d\mu _{B,R,m}(z),
\end{equation}%
with respect to the measure 
\begin{equation}
d\mu _{B,R,m}(z):=\frac{(2(BR^{2}-m)-1)}{\pi R^{2}\left( 1-\frac{z\bar{z}}{%
R^{2}}\right) ^{2}}d\mu (z),  \label{dmuBRm}
\end{equation}%
$d\mu (z)$ being the Lebesgue measure on the unit disk $\mathbb{D}$.
Recalling that the GCS $\left\vert \widetilde{\kappa }_{z,B,R,m}\right%
\rangle $ originate from Eq.~\eqref{2.13} according to a group theoretical
approach \cite{Perelomov}, it becomes natural to seek for a \textit{number
states} \textit{expansion} \cite{Dodonov} for them. Indeed, we can show that
they decompose as 
\begin{equation}
\left\vert \widetilde{\kappa }_{z,B,R,m}\right\rangle =\left( \mathcal{N}%
_{B,R,m}(z)\right) ^{-\frac{1}{2}}\sum_{j=0}^{+\infty }\gamma
_{j}^{B,R,m}(z)\left\vert j\right\rangle _{B,R,m}  \label{2.20}
\end{equation}%
where 
\begin{equation}
\mathcal{N}_{B,R,m}\left( z\right) :=\pi ^{-1}(2(BR^{2}-m)-1)\left( 1-z\bar{z%
}R^{-2}\right) ^{-2BR^{2}},
\end{equation}%
and the wavefunctions of the number states $\left|j\right\rangle_{B,R,m}$
are given by 
\begin{equation}
\langle \xi |j\rangle _{B,R,m}:=\left( \frac{2j!}{\Gamma (2(BR^{2}-m)+j)}%
\right) ^{\frac{1}{2}}\xi ^{2(BR^{2}-m)-\frac{1}{2}}e^{-\frac{\xi ^{2}}{2}%
}L_{j}^{(2(BR^{2}-m)-1)}(\xi ^{2}),\qquad\xi \in \mathbb{R}_{+},\text{ }
\label{2.22}
\end{equation}%
form an orthonormal basis in $L^{2}(\mathbb{R}_{+},d\xi)$. The coefficients 
\begin{equation}
\begin{split}
z\mapsto \gamma_{j}^{B,R,m}\left(z\right) := & \, (-1)^{m} \left( \frac{%
(2(BR^2-m)-1)}{\pi} \frac{(m\land j)!\Gamma(2(BR^2-m)+m\lor j)}{(m\lor
j)!\Gamma(2(BR^2-m)+m\land j)}\right)^{\frac{1}{2}} \\
& \frac{(-1)^{m\land j}}{\left(1-\frac{z\bar{z}}{R^2}\right)^{m}} \left(%
\frac{z\bar{z}}{R^2}\right)^{\frac{1}{2}|m-j|} e^{-i(m-j)\arg z} P_{m\land
j}^{(|m-j|, 2(BR^2-m)-1)}\left(1-\frac{2z\bar{z}}{R^2}\right)
\end{split}
\label{2.23}
\end{equation}
are in fact orthonormalized functions of the Hilbert space $L^{2}\left( 
\mathbb{D}_{R},R^{-2}\left( 1-\frac{z\bar{z}}{R^{2}}\right) ^{2BR^{2}-2}d\mu
(z)\right)$. Here $m\wedge j:=\min\{m,j\}$ and $m\vee j:=\max\{m,j\}$. The
key ingredient in proving Eq.~\eqref{2.20} is the use of the generating
formula (\cite[p.137]{Srivastava3}): 
\begin{equation}  \label{generating2F1Laguerre}
\sum_{n=0}^{+\infty }\lambda ^{n}{}_{2}\digamma_{1}\left( {\QATOP{-n,b}{1+a }%
}\Big|t\right) L_{n}^{(a)}(x)=\frac{(1-\lambda )^{b-1-a}}{ (1-\lambda
+t\lambda )^{b}}\exp \left( \frac{-x\lambda }{1-\lambda }\right)
{}_{1}\digamma_{1}\left( {\QATOP{b}{1+a }}\Big|\frac{\lambda xt}{(1-\lambda
)(1-\lambda +t\lambda )}\right)
\end{equation}
where the parameters $a \geq 0$ and $|\lambda |<\min \{1,|1-t|^{-1}\}$.

We may note in passing that for $R=1$, the GCS $\left\vert\widetilde{\kappa }%
_{z,B,1,m}\right \rangle$ in Eq.~\eqref{2.20} can be considered as a class of
generalized negative binomial states \cite{Mouayn5} and the above functions $%
\gamma_{j}^{B,1,m}(z)$ constitute, in fact, an orthonormal basis for the
eigenspace 
\begin{equation}
\mathcal{E}_{B,m}\left( \mathbb{D}\right) =\left\{ f \in L^{2}\left( \mathbb{%
D}, \left( 1-z\bar{z}\right) ^{2B-2}d\mu (z)\right), \quad\widetilde{\Delta}%
_{B} f =\sigma _{B,m} f \right\}  \label{2.28}
\end{equation}
of the $B$-weight Maass-Laplacian 
\begin{equation}
\widetilde{\Delta }_{B} = -4\left( 1-z\overline{z}\right) \left( \left( 1- z%
\overline{z}\right) \frac{\partial^{2}}{\partial z\partial \overline{z}}- 2B%
\overline{z}\frac{\partial}{\partial \overline{z}}\right)
\end{equation}%
associated with eigenvalues $\sigma _{B,m}=4m\left( 2B-m-1\right)$.
Additionally, the mapping%
\begin{equation}
W_{B,m}\left[ \varphi \right] \left( z\right) =\left( \mathcal{N}%
_{B,1,m}\left( z\right) \right) ^{\frac{1}{2}}\langle \varphi \left\vert 
\widetilde{\kappa }_{z,B,1,m}\right\rangle_{L^{2}(\mathbb{R}_{+},d\xi)}
\label{2.30}
\end{equation}%
is a unitary isomorphism from $L^{2}(\mathbb{R}_{+},d\xi )$ onto the Hilbert
space $\mathcal{E}_{B,m}\left( \mathbb{D}\right) $ whose reproducing kernel
can be obtained from the overlapping function $\langle \widetilde{\kappa }%
_{z,B,1,m},\, \widetilde{\kappa }_{w,B,1,m}\rangle _{L^{2}( \mathbb{R}%
_{+},d\xi)}$ between two GCS. Explicitly, 
\begin{equation}
\begin{split}
K_{B,m}\left( z,w\right) = & \, \pi ^{-1}(2\left( B-m\right) -1) \left( 1-z%
\overline{w}\right) ^{-2B}\left( \frac{\left\vert 1-z\overline{w}
\right\vert^{2}}{\left(1-z\overline{z}\right) \left(1-w\overline{w}\right)}
\right)^{m} \\
& \times P_{m}^{\left( 0,2\left( B-m\right) -1\right) } \left( 2\frac{\left(
1-z\overline{z}\right) \left( 1-w\overline{w}\right) } {\left\vert 1-z%
\overline{w}\right\vert^{2}}-1\right)
\end{split}
\label{KBmzw}
\end{equation}
for $z,w\in\mathbb{D}$.

\section{Husimi's Q-function for a pure state}

\label{section4}

By using Eq.~\eqref{2.20} to Eq.~\eqref{2.23}, one can write according to Eq.~%
\eqref{1.2}, the Husimi function of the pure state%
\begin{equation}
\widehat{\rho }_{j}:=|j\rangle _{B,R,m\text{ \ }B,R,m}\langle j|,
\label{3.1}
\end{equation}%
as 
\begin{equation}
Q_{j}^{B,R,m}(z):=\langle \widetilde{\kappa }_{z,B,R,m}|\widehat{\rho }_{j}|%
\widetilde{\kappa }_{z,B,R,m}\rangle ,\qquad z\in \mathbb{D}_{R}.  \label{3.1}
\end{equation}%
Explicitly, 
\begin{equation}
Q_{j}^{B,R,m}(z)=\tau _{j}^{B,R,m}\left( 1-\frac{z\bar{z}}{R^{2}}\right)
^{2(BR^{2}-m)}\left( \frac{z\bar{z}}{R^{2}}\right) ^{|m-j|}\left( P_{m\wedge
j}^{(|m-j|,2(BR^{2}-m)-1)}\left( 1-\frac{2z\bar{z}}{R^{2}}\right) \right)
^{2}  \label{3.4}
\end{equation}%
where 
\begin{equation}
\tau _{j}^{B,R,m}:=\frac{(m\wedge j)!\Gamma (2(BR^{2}-m)+m\vee j)}{(m\vee
j)!\Gamma (2(BR^{2}-m)+m\wedge j)}  \label{Tauj}
\end{equation}%
for every fixed $m=0,1,\dots ,\lfloor BR^{2}-\frac{1}{2}\rfloor $ provided
that $2BR^{2}>1$. 
Doing so enables us, by letting $R\to\infty$, to recover the (radial)
Husimi's Q-function for a pure state associated with an $m$th Euclidean
Landau level (\cite[p.5]{Mouayn7}): 
\begin{equation}
Q_{j}^{(m),\text{ Euclid}}(\sqrt{2B}z)=\frac{(m\wedge j)!}{(m\vee j)!}
e^{-2Bz\bar{z}}\left( 2Bz\bar{z}\right)^{|m-j|} \left( L_{m\wedge
j}^{(|m-j|)} (2Bz\bar{z})\right) ^{2},\; \qquad m=0,1,2,\dots ,\quad z\in \mathbb{C},
\label{3.6}
\end{equation}
see Appendix \ref{AppB} for the proof. We also note that the characteristic
function for the random variable $X$ having $Q_{j}^{(m),\text{ Euclid}}$ as
its density function reads (\cite[p.6]{Mouayn7}):
\begin{eqnarray}
\phi_{X}( u) &:=& \left( 1-iu\right)^{-(m+j+1)}
\sum_{k=0}^{m\land j} \binom{m}{k} \binom{j}{k} (iu)^{2k},\qquad u\in \mathbb{R} \label{CFPureEuclidean0} \\
&=&\frac{(m+j)!}{m!j!}\left( 1-iu\right)
^{-(m+j+1)}{}_{2}\digamma_{1}\left( {\QATOP{-m,-j}{-m-j}}\Big|1+u^{2}\right) \notag
\end{eqnarray}
and was involved by Shirai while discussing Ginibre-type processes \cite%
{Shirai}. 
Like the Euclidean setting, one can check that the Husimi function given by
Eq.~\eqref{3.4} is a probability distribution on the phase space $\mathbb{D}%
_{R}$. That is, 
\begin{equation}
\int\limits_{\mathbb{D}_{R}}Q_{j}^{B,R,m}(z)d\mu _{B,R,m}(z)=1  \label{3.8}
\end{equation}%
where $d\mu _{B,R,m}(z)$ is given by Eq.~\eqref{dmuBRm}. Using polar
coordinates $z=re^{i\theta }$,\quad $r\in \lbrack 0,R)$, \quad $\theta \in \left[
0,2\pi \right) $ and setting 
\begin{equation}
\mathfrak{Q}_{j}^{B,R,m}(\lambda):= R^{-2}\left( 2(BR^{2}-m)-1\right)
\left(1-\lambda R^{-2}\right)^{-2} {Q}_{j}^{B,R,m}(\lambda ),  \label{3.10}
\end{equation}%
with $\lambda =r^{2},$ then Eq.~\eqref{3.8} ensures 
\begin{equation}
\int\limits_{0}^{R^{2}}\mathfrak{Q}_{j}^{B,R,m}(\lambda )d\lambda =1.
\label{3.11}
\end{equation}
For $m=0$, the density function in Eq.~\eqref{3.10} reduces to 
\begin{equation}
\mathfrak{Q}_{j}^{B,R,0}(\lambda) = R^{-2} \frac{\Gamma (2BR^{2}-1+j+1)}{%
\Gamma (2BR^{2}-1)\Gamma (j+1)} \left( 1-\frac{\lambda }{R^{2}}%
\right)^{(2BR^{2}-1)-1} \left( \frac{\lambda }{R^{2}}\right) ^{(j+1)-1}
\label{HusimiPurem0}
\end{equation}%
which coincides (up to a scale factor $1/R^{2}$) with the Beta distribution $%
\mathcal{B}e(j+1,2BR^{2}-1)$ whose characteristic function is known to be
given by the confluent hypergeometric series as 
\begin{equation}
\phi _{\mathcal{B}e}(u)={}_{1}\digamma_{1}\left( {\QATOP{{j+1}}{{2BR^{2}+j}}}%
\bigg|iuR^{2}\right).  \label{CFPurem0}
\end{equation}%
\ \newline
If for $R=1$ we denote by $\mathfrak{X}_{m}^{\left( 1\right) }$ the random
variable having $\lambda \mapsto \mathfrak{Q}_{j}^{B,1,m}(\lambda) $ as its
density function then $\mathfrak{X}_{m}^{(1)}$ would play the role of the
hyperbolic analog of the continuous random variable denoted by $S_{k}^{n,b}$
in Shirai's paper \cite{Shirai} and 
\begin{equation}
\Pr \left( \mathfrak{X}^{(1)}_{m}\leq r^{2}\right) =\int\limits_{0}^{r^{2}}%
\mathfrak{Q}_{j}^{B,1,m}(\xi)d\xi \equiv \lambda _{j}^{m,B}( r^{2})
\end{equation}
would provide a probabilistic representation of eigenvalues $\lambda
_{j}^{m,B}(r^{2}):=||\gamma_{j}^{B,1,m} \mathbf{1}_{\mathbf{D}_r}||^2$ of
the restricted operator $\left(K_{B,m}( z,w)\right)_{\mathbf{D}_{r}}$ to the
disk $\mathbf{D}_{r}\subset \mathbb{D}$ of radius $r\leq 1,$ where $%
K_{B,m}\left( z,w\right)$ defined in Eq.~\eqref{KBmzw} is the projection
operator onto the $m$th hyperbolic Landau eigenspace $\mathcal{E}_{B,m}
\left( \mathbb{D}\right)$ in Eq.~\eqref{2.28}. Furthermore, the operator $%
\left( K_{B,m}\left( z,w\right)\right)_{\mathbf{D}_{r}}$ may be unitarly
intertwined via the Bargmann-type transform $W_{B,m}$ in Eq.~\eqref{2.30} to
be acting on the quantum (or signals) Hilbert space $L^{2}(\mathbb{R}%
_{+},d\xi )$ as a localization operator \textquotedblleft \textit{\`{a} la
Daubechies}\textquotedblright \cite{Daubechies} of the indicator function $%
\mathbf{1}_{\mathbf{D}_r}$ of the disk $\mathbf{D}_{r}$.

For these reasons, it would be useful to investigate properties of {the
random variable }$\mathfrak{X}_{m}^{(R)}${\ having }$\mathfrak{Q}%
_{j}^{B,R,m}(\lambda)$ as its density function by first writing down its
characteristic function which is defined, as usual, by the Fourier-type
integral transform (\cite[p.22]{CF}):%
\begin{equation}
\phi _{\mathfrak{X}_{m}^{\left( R\right) }}\left(
u\right):=\int\limits_{0}^{R^{2}} e^{iu\lambda }\mathfrak{Q}%
_{j}^{B,R,m}(\lambda) d\lambda ,\quad u\in \mathbb{R}.  \label{CFPureDef}
\end{equation}
Straightforward calculations (see Appendix \ref{AppC}) lead to the expression%
\begin{equation*}
\phi _{\mathfrak{X}_{m}^{\left( R\right) }}\left( u\right) =\tau
_{j}^{B,R,m}(2(BR^{2}-m)-1)\sum_{k=0}^{2(m\wedge j)}C_{j,k}^{B,R,m}\Gamma
(2(BR^{2}-m)-1+k)\frac{\left( -iuR^{2}\right) ^{k}}{k!}
\end{equation*}%
\begin{equation}
\times \frac{\Gamma \left( \left\vert m-j\right\vert +k+1\right) }{\Gamma
\left( 2(BR^{2}-m)+\left\vert m-j\right\vert +2k\right) }._{1}\digamma
_{1}\left( 
\begin{array}{c}
\left\vert m-j\right\vert +k+1 \\ 
2(BR^{2}-m)+\left\vert m-j\right\vert +2k%
\end{array}%
\mid iuR^{2}\right)  \label{CFPure}
\end{equation}
where 
\begin{equation*}
C_{j,k}^{B,R,m}=\frac{(2(m\wedge j))!(|m-j|+1))_{2(m\wedge
j)}((2(BR^{2}-m)+|m-j|)_{2(m\wedge j)})^{2}}{((m\wedge
j)!)^{2}((2(BR^{2}-m)+|m-j|)_{m\wedge j})^{2}}
\end{equation*}%
\begin{equation*}
\times \frac{(-1)^{k}(2(BR^{2}-m)+|m-j|-1)_{k}(2k+2(BR^{2}-m)+|m-j|-1)}{%
(2(m\wedge j)-k)!(|m-j|+1)_{k}(2(BR^{2}-m)+|m-j|-1)_{2(m\wedge j)+k+1}}
\end{equation*}
\begin{equation}
\times \digamma _{2:1:1}^{2:2:2}\left( 
\begin{array}{c}
-2(m\wedge j)+k,-2(BR^{2}-m)-m-j+1-k;-m,-j;-m,-j \\ 
-2(m\wedge j),-m-j;-2(BR^{2}-m)-m-j+1;-2(BR^{2}-m)-m-j+1%
\end{array}%
\bigg|1,1\right).  \label{Cjk}
\end{equation}
Here, $\digamma _{2:2:2}^{2:1:1}$ denotes the Kamp\'{e} de F\'{e}riet
function of two variables defined by Eq.~\eqref{kampe}. For $m=0$, Eq.~%
\eqref{CFPure} reduces to Eq.~\eqref{CFPurem0} as expected.

Now, by having $\phi _{\mathfrak{X}_{m}^{\left( R\right) }}\left( u\right)$,
we immediately derive the main statistical parameters which are the mean
value and the variance of {$\mathfrak{X}_{m}^{\left( R\right) }$}. Namely,
we obtain the following formulas 
\begin{equation}
\mathbb{E}(\mathfrak{X}_{m}^{\left( R\right) })=\tau _{j}^{B,R,m}R^{2}\frac{%
(|m-j|+1)!\Gamma (2(BR^{2}-m))}{\Gamma (2(BR^{2}-m)+|m-j|+1)}\left(
C_{j,0}^{B,R,m}-\frac{(2(BR^{2}-m)-1)C_{j,1}^{B,R,m}}{(2(BR^{2}-m)+|m-j|+1)}%
\right)  \label{EsperenceHusimiPure}
\end{equation}%
and 
\begin{equation}
\mathbb{V}ar\left( \mathfrak{X}_{m}^{\left( R\right) }\right) = R^4 \tau
_{j}^{B,R,m}\frac{(|m-j|+2)!\Gamma (2(BR^{2}-m))}{\Gamma
(2(BR^{2}-m)+|m-j|+2)}  \label{VarHusimiPure}
\end{equation}%
\begin{equation*}
\bigg(C_{j,0}^{B,R,m}-\frac{2((2(BR^{2}-m)-1))C_{j,1}^{B,R,m}}{%
(2(BR^{2}-m)+|m-j|+2)}+\frac{(2(BR^{2}-m)-1)(2(BR^{2}-m))C_{j,2}^{B,R,m}}{%
(2(BR^{2}-m)+|m-j|+2)(2(BR^{2}-m)+|m-j|+3)}\bigg)
\end{equation*}%
\begin{equation*}
-\bigg[R^2\tau _{j}^{B,R,m}\frac{(|m-j|+1)!\Gamma (2(BR^{2}-m))}{\Gamma
(2(BR^{2}-m)+|m-j|+1)}\left( C_{j,0}^{B,R,m}-\frac{%
(2(BR^{2}-m)-1)C_{j,1}^{B,R,m}}{(2(BR^{2}-m)+|m-j|+1)}\right)\bigg]^{2}
\end{equation*}%
For the proofs of Eq.~\eqref{EsperenceHusimiPure} and Eq.~\eqref{VarHusimiPure}
see Appendix \ref{AppE}. In particular, for $m=0$ the above formulas reduce
to

\begin{equation}
\mathbb{E}\left( \mathfrak{X}_{0}^{\left( R\right) }\right) =\frac{(j+1)R^{2}%
}{2BR^{2}+j},\quad \mathbb{V}ar(\mathfrak{X}_{0}^{\left( R\right) })=\frac{%
(j+1)(2BR^{2}-1)R^{4}}{(2BR^{2}+j)^{2}(2BR^{2}+j+1)},  \label{3.18}
\end{equation}%
which are (up to a scale factor $1/R^{2}$) the mean value and variance of
the Beta distribution $\mathcal{B}e\left( j+1,2BR^{2}-1\right)$. Finally, we
establish (see Appendix \ref{AppD}) the following limit

\begin{equation}
\phi _{\mathfrak{X}_{m}^{\left( R\right) }}\left( u\right) \rightarrow
\left( 1-\frac{iu}{2B}\right)^{-(m+j+1)} \sum_{k=0}^{m\land j} \binom{m}{k} 
\binom{j}{k} \left(\frac{iu}{2B}\right)^{2k} ,\qquad u<2B
\label{CFPureConvergence}
\end{equation}
as $R\rightarrow +\infty$, which turn out to be (up to a scale factor $1/2B$%
) the characteristic function in the corresponding Euclidean setting in Eq.~%
\eqref{CFPureEuclidean0}. 
\newline
\newline
\textbf{Remark 1.} Note that ${Q}_{j}^{B,R,m}(z)$ vanishes at the origin $%
(0,0)$, the points of the circle of radius $R$ and the points $z$ such that
that $R^{2}-2\left\vert z\right\vert ^{2}$ are zeros of the Jacobi
polynomials $P_{m\wedge j}^{(|m-j|,2(B-m)-1)}(\cdot )$. Denoting by $%
x_{i}^{(m\wedge j)},\quad 1\leq i\leq m\wedge j$ the ordered consecutive zeros of
this polynomial, the zeros of ${Q}_{j}^{B,R,m}(z)$ are then located on the
concentric circles of radius $r_{i}=\sqrt{\left( R^{2}-x_{i}^{(m\wedge
j)}\right) /2}$. A discussion on the number of zeros of the Husimi density
and their location in connection with properties of harmonic and anaharmonic
oscillators can be found in \cite{Korsh}, {particularly \cite{Mouayn7}} in
the Euclidean setting.\ \newline

\section{Husimi's Q-function for mixed states}

\label{section5}

\subsection{A Hamiltonian operator for the GCS}

Recall that the functions $\left\langle\xi|j\right\rangle_{B,R,m}$ in the
expansion \eqref{2.20} of the GCS $\left\vert \widetilde{\kappa }%
_{z,B,R,m}\right\rangle $ were given by the normalized Laguerre functions %
\eqref{2.22} in $L^{2}(\mathbb{R}_{+},d\xi )$. Moreover, one may use the
Rodriguez formula for the Laguerre polynomial to show that these functions
can also be obtained by a $j$-fold application of the first order
differential operator

\begin{equation}
A_{\tau ,m}=\frac{1}{\sqrt{2}}\left( -\frac{\tau-m-1/2}{\xi }+\xi +\frac{d}{%
d\xi }\right) ,\qquad \tau =BR^{2},
\end{equation}%
on the ground state 
\begin{equation}
\langle \xi |0\rangle _{B,R,m}:=\left( \frac{2}{\Gamma (2(BR^{2}-m))}\right)
^{\frac{1}{2}}\xi ^{2(BR^{2}-m)-\frac{1}{2}}e^{-\frac{\xi ^{2}}{2}},\qquad \xi \in \mathbb{R}_{+},
\end{equation}%
as follows 
\begin{equation}
|j\rangle _{B,R,m}=\left( A_{\tau ,m}\right) ^{j}|0\rangle _{B,R,m}.
\end{equation}%
The formal adjoint operator of $A_{\tau ,m}$ in $L^{2}(\mathbb{R}_{+},d\xi)$
is given by 
\begin{equation}
A_{\tau ,m}^{\ast }=\frac{1}{\sqrt{2}}\left( -\frac{\tau-m-1/2}{\xi }+\xi - 
\frac{d}{d\xi }\right).
\end{equation}%
Therefore we may associate to GCS $\left\vert \widetilde{\kappa }%
_{z,B,R,m}\right\rangle $ the Hamiltonian operator 
\begin{equation}
H_{\tau ,m}=A_{\tau ,m}^{\ast }A_{\tau ,m}+2\left( \tau-m\right).
\end{equation}%
Explicitly, 
\begin{equation}
H_{\tau ,m}:=-\frac{1}{2}\frac{d^{2}}{d\xi ^{2}}+\frac{\left(
2(\tau-m)-1\right) ^{2}-1/4}{2\xi ^{2}}+\frac{1}{2}\xi ^{2}  \label{Htaum}
\end{equation}
which appears in the literature under many names such as isotonic oscillator 
\cite{isotonic1,isotonic2}, Gol'dman-Krivchenkov Hamiltonian \cite{Goldman},
pseudoharmonic oscillator \cite{Popov} or Laguerre operator \cite{Bentacor},
see also (\cite[p.171]{Mourad}). 
The spectrum of $H_{\tau ,m}$ in $L^{2}(\mathbb{R}_{+},d\xi)$ is purely
discrete and given by the eigenvalues 
\begin{equation}
\eta _{j}=j+4(\tau-m)-1.
\end{equation}%
The latter ones, together with their associated eigenfunctions $|j\rangle
_{B,R,m}$ can be used to define the heat semigroup associated with $%
H_{\tau,m}$ through the discret spectral resolution \textit{\ } 
\begin{equation}
e^{-\beta H_{\tau ,m}}\left[ f\right] :=\sum_{j=0}^{+\infty }e^{-\beta \eta
_{j}}\left\langle f\right\vert j\rangle _{B,R,m}.\left\vert j\right\rangle
_{B,R,m}, \;\; f\in L^{2}(\mathbb{R}_{+}).  \label{4.8}
\end{equation}%
It is also well known that by using the Hille-Hardy formula (\cite[p.242]%
{Magnus}), this semigroup has the integral representation 
\begin{equation}
e^{-\beta H_{\tau ,m}}\left[ f\right] \left( r\right)
=\int\limits_{0}^{+\infty }W_{\beta }^{B,m}\left( r,\rho \right) f\left(
\rho \right) d\rho
\end{equation}%
where 
\begin{equation}
W_{\beta }^{B,m}\left( r,\rho \right) = \frac{2\sqrt{r\rho }e^{-\beta }} {%
(1-e^{-2\beta })} \exp \left( -\frac{1}{2}\left( r^{2}+\rho ^{2}\right) 
\frac{1+e^{-2\beta }}{1-e^{-2\beta }}\right) I_{\left( 2(BR^{2}-m)-1\right)
}\left( \frac{2r\rho e^{-\beta }}{1-e^{-2\beta }}\right) \text{.}
\end{equation}%
Here $I_{a}\left( \cdot \right) $ denotes the modified Bessel function
defined in Eq.~\eqref{IBessel}.

\subsection{Husimi's function}

The standard statistical mechanics starts using the Gibb's canonical
distribution, whose thermal density operator is represented by 
\begin{equation}
\hat{\rho}_{\beta }=\frac{1}{Z}e^{-\beta \hat{H}}
\end{equation}%
where $Z=\mathrm{Tr}\left( e^{-\beta \hat{H}}\right) $ is the partition function, $%
\hat{H}$ is the Hamiltonian of the system, $\beta =1/(kT)$ the inverse
temperature $T$ and $k$ the Boltzmann constant ($k=1$). In our context, $%
\hat{H}\equiv H_{\tau ,m}$ the isotonic oscillator and $\hat{\rho}_{\beta }$
is the associated normalized heat operator. The Husimi distribution is
defined as the expectation value of the density operator $\hat{\rho}_{\beta }
$ in the set of states $|\widetilde{\kappa }_{z,B,R,m}\rangle $ as 
\begin{equation}
Q_{\beta }^{B,R,m}(z):=\langle \widetilde{\kappa }_{z,B,R,m}|\widehat{\rho }%
_{\beta }|\widetilde{\kappa }_{z,B,R,m}\rangle ,\qquad z\in \mathbb{D}_{R}.
\label{QMixedDef}
\end{equation}%
By using Eq.~\eqref{4.8}, the R.H.S of Eq.~\eqref{QMixedDef} takes the form 
\begin{equation}
Q_{\beta }^{B,R,m}(z)=\frac{1}{Z}\sum_{j=0}^{+\infty }e^{-\beta \eta _{j}}%
\left[ Q_{j}^{B,R,m}(z)\right] 
\end{equation}%
where $Q_{j}^{B,R,m}$ is the Husimi function for the pure state $|j\rangle
\langle j|$ given by Eq.~\eqref{3.4}, where $|j\rangle \equiv |j\rangle
_{B,R,m}$. Replacing the $\eta _{j}$ by their expressions $j+4(BR^{2}-m)-1$
and using direct calculations, we obtain the partition function as 
\begin{equation}
Z=\frac{1}{1-e^{-\beta }}e^{-(4(BR^{2}-m)-1)\beta }
\end{equation}%
such that 
\begin{equation}
\int\limits_{\mathbb{D}_{R}}Q_{\beta }^{B,R,m}(z)d\mu _{B,R,m}\left(
z\right) =1.  \label{RIMS}
\end{equation}%
Therefore, 
\begin{equation}
Q_{\beta }^{B,R,m}(z)=\left( 1-e^{-\beta }\right) \sum_{j=0}^{+\infty
}\left( e^{-\beta }\right) ^{j}\left[ Q_{j}^{B,R,m}(z)\right] .
\label{Qrhot_pr3}
\end{equation}%
Next, if we make the last sum look like the moment generating function of
the generalized negative binomial distribution (\cite[p.6]{Mouayn6}), then
we get 
\begin{equation}
\begin{split}
Q_{\beta }^{B,R,m}(z)=& \left( 1-e^{-\beta }\right) \left( \frac{1-\frac{z%
\bar{z}}{R^{2}}}{1-\frac{z\bar{z}}{R^{2}}e^{-\beta }}\right)
^{2BR^{2}}\left( \frac{\left( \frac{z\bar{z}}{R^{2}}-e^{-\beta }\right)
\left( 1-\frac{z\bar{z}}{R^{2}}e^{-\beta }\right) }{\left( 1-\frac{z\bar{z}}{%
R^{2}}\right) ^{2}}\right) ^{m} \\
& \times P_{m}^{(2(BR^{2}-m)-1,0)}\left( 1+\frac{2e^{-\beta }\left( 1-\frac{z%
\bar{z}}{R^{2}}\right) ^{2}}{\left( \frac{z\bar{z}}{R^{2}}-e^{-\beta
}\right) \left( 1-\frac{z\bar{z}}{R^{2}}e^{-\beta }\right) }\right) .
\end{split}%
\end{equation}%
By letting $R\rightarrow \infty $ we prove (see Appendix \ref{AppF}) that $%
Q_{\beta }^{B,R,m}(z)\rightarrow q_{m}(|\sqrt{2B}z|^{2},\beta )$ where 
\begin{equation}
q_{m}(|\sqrt{2B}z|^{2},\beta )=\left( 1-e^{-\beta }\right) \exp \left( -2|%
\sqrt{2B}z|^{2}(1-e^{-\beta })\right) e^{-\beta m}L_{m}^{(0)}\left( -4|\sqrt{%
2B}z|^{2}\sinh ^{2}\frac{\beta }{2}\right)   \label{QMixedConvergence}
\end{equation}%
is the Husimi function for mixed states of the flat case (\cite[p.7]{Mouayn7}%
) as expected. Now, setting 
\begin{equation}
\mathfrak{Q}_{\beta }^{B,R,m}(\lambda ):=R^{-2}\left( 2(BR^{2}-m)-1\right)
\left( 1-\lambda R^{-2}\right) ^{-2}{Q}_{\beta }^{B,R,m}(\lambda ),
\label{QMixedl}
\end{equation}%
with $\lambda =z\bar{z}$, Eq.~\eqref{RIMS} gives 
\begin{equation}
\int\limits_{0}^{R^{2}}\mathfrak{Q}_{\beta }^{B,R,m}(\lambda )d\lambda =1.
\end{equation}%
Then, we denote by $\mathfrak{Y}_{m}^{\left( R\right) }$ the random variable
having $\lambda \mapsto \mathfrak{Q}_{\beta }^{B,R,m}(\lambda )$ as its
density function for which it would be useful to investigate some properties
by starting by its characteristic function which is defined, as usual, by 
\begin{equation}
\phi _{\mathfrak{Y}_{m}^{\left( R\right)
}}(u):=\int\limits_{0}^{R^{2}}e^{iu\lambda }\mathfrak{Q}_{\beta
}^{B,R,m}(\lambda )d\lambda ,\qquad u\in \mathbb{R}.  \label{4.22}
\end{equation}%
Straightforward calculations (see Appendix \ref{CFQMixedP}) lead to the
expression 
\begin{equation}
\begin{split}
\phi _{\mathfrak{Y}_{m}^{\left( R\right) }}(u)=& (2(BR^{2}-m)-1)(1-e^{-\beta
})e^{-m\beta }\sum_{k=0}^{m}\binom{2BR^{2}-m-1}{k}\binom{m}{k}\left( \frac{%
(1-e^{-\beta })^{2}}{e^{-\beta }}\right) ^{k} \\
& \mathcal{B}(k+1,2BR^{2}-2k-1)\,\Phi _{1}\left( {\QATOP{k+1,2BR^{2}}{%
2BR^{2}-k}}\Big|e^{-\beta },-iuR^{2}\right) 
\end{split}
\label{CFQMixed}
\end{equation}%
where $\mathcal{B}(\cdot ,\cdot )$ is the beta function and $\Phi _{1}$ is
the Humbert series defined in Eq.~%
\eqref{Phi1Def}.\newline
Now, by having $\phi _{\mathfrak{Y}_{m}^{\left( R\right) }}\left( u\right) $%
, we immediately derive (see Appendix \ref{EVQMixedP} for the proof) the
mean value and the variance of $\mathfrak{Y}_{m}^{\left( R\right) }$ that
are given respectively by 
\begin{equation}
\begin{split}
\mathbb{E}\left( \mathfrak{Y}_{m}^{\left( R\right) }\right) =&
\;R^{2}(2(BR^{2}-m)-1)(e^{-\beta }-1)e^{-m\beta }\sum_{k=0}^{m}\binom{%
2BR^{2}-m-1}{k}\binom{m}{k}\left( \frac{(1-e^{-\beta })^{2}}{e^{-\beta }}%
\right) ^{k} \\
& \mathcal{B}(k+2,2BR^{2}-2k-1)\,{}_{2}\digamma _{1}\left( {\QATOP{%
k+2,2BR^{2}}{2BR^{2}-k+1}}\Big|e^{-\beta }\right) 
\end{split}
\label{EQMixed}
\end{equation}%
and 
\begin{equation*}
\mathbb{V}ar(\mathfrak{Y}_{m}^{\left( R\right)
})=R^{4}(2(BR^{2}-m)-1)(1-e^{-\beta })e^{-m\beta }
\end{equation*}%
\begin{equation}
\begin{split}
\times & \sum_{k=0}^{m}\binom{2BR^{2}-m-1}{k}\binom{m}{k}\left( \frac{%
(1-e^{-\beta })^{2}}{e^{-\beta }}\right) ^{k}\mathcal{B}%
(k+3,2BR^{2}-2k-1){}_{2}\digamma _{1}\left( {\QATOP{k+3,2BR^{2}}{2BR^{2}-k+2}%
}\Big|e^{-\beta }\right)  \\
-& \Bigg[R^{2}(2(BR^{2}-m)-1)(1-e^{-\beta })e^{-m\beta }\sum_{k=0}^{m}\binom{%
2BR^{2}-m-1}{k}\binom{m}{k}\left( \frac{(1-e^{-\beta })^{2}}{e^{-\beta }}%
\right) ^{k} \\
\times & \mathcal{B}(k+2,2BR^{2}-2k-1){}_{2}\digamma _{1}\left( {\QATOP{%
k+2,2BR^{2}}{2BR^{2}-k+1}}\Big|e^{-\beta }\right) \Bigg]^{2}.
\end{split}
\label{VQMixed}
\end{equation}

We also establish (see Appendix \ref{LimitCFQMixedP}) the following limit 
\begin{equation}
\phi_{\mathfrak{Y}_{m}^{\left(R\right)}}( u) \to \frac{(1-e^{-\beta})} {%
1-e^{-\beta}-\frac{iu}{2B}} \left(\frac{1-e^{-\beta}-\frac{iu}{2B}e^{-\beta}%
} {1-e^{-\beta}-\frac{iu}{2B}} \right)^m,\qquad u<2B  \label{LimitCFQMixed}
\end{equation}
as $R\to+\infty$, which turn out to be (up to a scale factor $\frac{1}{2B}$)
the characteristic function in the Euclidean setting \cite[p.8]{Mouayn7} as
expected. \newline
\newline
\textbf{Remark 2.} For $q_m(|\sqrt{2B}z|^2,\beta)$ in Eq.~%
\eqref{QMixedConvergence}, we may set $\lambda =2Bz\overline{z}$ and use a
duality to define a non negative integer-valued random variable $\mathcal{Y}%
_{\lambda,\beta }$ by 
\begin{equation}
\Pr \left( \mathcal{Y}_{\lambda ,\beta }\right) =q_{m}\left( \lambda ,\beta
\right) ,\qquad m=0,1,2,...,
\end{equation}
to recover the Laguerre probability distribution with parameters $\left(
\lambda ,\beta \right) $ which can be presented as follows. Let $T=1/k\beta $
represents in suitable units $(k=1)$, the absolute temperature and $%
N_{T}:=\left( e^{\frac{1}{T}}-1\right) ^{-1}$ the average number of photons
associated with the thermal noisy state, then the photon count probability
distribution for a mixed light reads (\cite{Perina}): 
\begin{equation}
q_{m}\left( \lambda ,\beta \right) =\frac{(N_{T})^{m}}{(1+N_{T})^{m+1}}\exp
\left( -\frac{{\lambda }}{1+N_{T}}\right) L_{m}^{(0)}\left( -\frac{\lambda }{
N_{T}(1+N_{T})}\right) .
\end{equation}
\newline
\newline
\indent
Finally, we can use the obtained Q-function $Q_{\beta}^{B,R,m}$ to get a
lower bound for the thermodynamical potential associated with the
Hamiltonian $H_{\tau,m}$, $\tau=BR^2$. This potential reads 
\begin{equation}
\Theta_{\beta}^{\tau,\eta} := \frac{-1}{\beta} \mathrm{Tr}\left(\log\left(
1+e^{-\beta(H_{\tau,m}-\eta)}\right)\right)  \label{ThermoDynPotential}
\end{equation}
where $\eta$ is the chemical potential. Putting $\epsilon=e^{\beta\eta}$,
the form of the potential in Eq.~\eqref{ThermoDynPotential} suggests us to
consider the function 
\begin{equation}
\zeta \mapsto \phi_{\epsilon}(\zeta) = - \log(1+\epsilon\zeta).
\label{phiespilon}
\end{equation}
So that we can rewrite Eq.~\eqref{ThermoDynPotential} as 
\begin{equation}
\beta \Theta_{\beta}^{\tau,\eta} = \mathrm{Tr}\left(\phi_{\epsilon}\left( e^{-\beta
H_{\tau,m}} \right)\right).  \label{BerezinP2}
\end{equation}
Applying the Berezin-Lieb inequality \cite{Berezin} for the lower symbol $%
Q_{\beta}^{B,R,m}$  of the operator $e^{-\beta H_{\tau ,m}}$ as defined by 
Eq.~\eqref{QMixedDef}, we obtain that 
\begin{equation}
\int\limits_{\mathbb{D}_R} [\phi_{\epsilon} \circ Q_{\beta}^{B,R,m} ](z)
d\mu_{B,R,m}(z) \leq \mathrm{Tr}(\phi_\epsilon(e^{-\beta H_{\tau,m}})).
\label{BerezinP1}
\end{equation}
Making use of Eq.~\eqref{phiespilon} and replacing the R.H.S of Eq.~%
\eqref{BerezinP1} by $\beta \Theta_{\beta}^{\tau,\eta}$ as in Eq.~%
\eqref{BerezinP2}, we get an inequality that holds for every $m=0,1
,\dots,\left\lfloor BR^2-1/2\right\rfloor$. Therefore, we consider the
maximum with respect to the integer $m$ of the quantity in the L.H.S of Eq.~%
\eqref{BerezinP1} as 
\begin{equation}
\max_{m\in\mathbb{Z}_+\cap[0,BR^2-1/2]} \left[\frac{1}{\beta} \int\limits_{%
\mathbb{D}_R} \log\left(\frac{1}{1+\epsilon Q_{\beta}^{B,R,m}(z)}\right)
d\mu_{B,R,m}(z) \right] \leq \Theta_{\beta}^{BR^2,\eta}
\end{equation}
for every $\beta>0$.

\section{A formula for a special Kamp\'e de F\'eriet function}

\label{section6} In this section we establish summation formula for the
Kamp\'e de F\'eriet function $\digamma_{2:0:0}^{1:2:2}$, wich will be limit
in Eq.~\eqref{3.6} (see appendix \ref{AppD}). \newline
\newline
\textbf{Proposition 1.} \textit{Let $n\in\mathbb{N}^*$. For all real $x>0$ or $x<-2\sqrt{2}$, the
equality} 
\begin{equation}
\sum_{k=0}^{n}\binom{n}{k}\digamma_{2:0:0}^{1:2:2}\left({\QATOP{%
-n+k;a,b;c-a,c-b}{-n,c;-;-}}\bigg|1,1\right) x^{k}=(1+x)^{n}\left( \frac{x}{%
1+x}\right) ^{a+b-c}{}_{2}\digamma_{1}\left( {\QATOP{a,b}{c}}\bigg|\frac{1+2x%
}{(1+x)^{2}}\right)  \label{FormuleKampe}
\end{equation}
\textit{holds true. It also holds true for $x\in\mathbb{C}\setminus\{-1,0\}$ if $%
\max\{a,b\}\in\mathbb{Z}_{-}$ and $\max\{c-a,c-b\}\in\mathbb{Z}_{-}$.} 
\newline
\textit{Proof.} We denote 
\begin{equation}
S_{n}:=\sum_{k=0}^{n}\binom{n}{k}\digamma_{2:0:0}^{1:2:2}\left( {\QATOP{%
-n+k;a,b;c-a,c-b}{-n,c;-;-}}\bigg|1,1\right) x^{k}.
\end{equation}
for $n\in\mathbb{N}^*$. Using the definition of $\digamma_{l:m:n}^{p:q:k}$ in Eq.~\eqref{kampe} we
obtain 
\begin{equation}
S_{n} = \sum_{s,t\geq 0}\frac{(a)_{s}(b)_{s}(c-a)_{t}(c-b)_{t}}{%
(-n)_{s+t}(c)_{s+t}}\frac{1}{s!t!}\sum_{k=0}^{n}\binom{n}{k}%
(-n+k)_{s+t}x^{k}.  \label{A.2}
\end{equation}
Since $(-n+k)_{s+t}=0$ if $s+t>n-k$, the second sum in the R.H.S of Eq.~\eqref{A.2} terminates as 
\begin{equation}
\sum_{k=0}^{n}\binom{n}{k}(-n+k)_{s+t}x^{k} = (-n)_{s+t} \sum_{k=0}^{n-s-t} 
\binom{n-s-t}{k} x^k = (-n)_{s+t}(1+x)^{n-s-t}  \label{A.3}
\end{equation}
when $(-n)_{k}=(-1)^{k}k!\binom{n}{k}$ for $0\leq k\leq n$, $n\in\mathbb{N}^*$. Therefore, 
Eq.~\eqref{A.2} takes the form 
\begin{equation}
S_{n} = (1+x)^{n}\sum_{s,t\geq 0}\frac{(a)_{s}(b)_{s}(c-a)_{t}(c-b)_{t}} {%
(c)_{s+t}}\frac{1}{s!t!}\left( \frac{1}{1+x}\right)^{s+t}  \label{Eq.6.5}
\end{equation}%
wich also reads 
\begin{equation}
S_{n} = (1+x)^{n}\sum_{s\geq 0}\frac{(c-a)_{s}(c-b)_{s}}{(c)_{s}}{}%
_{2}\digamma_{1}\left( {\QATOP{a,b}{c+s}}\bigg|\frac{1}{1+x}\right) \frac{1}{%
s!} \left( \frac{1}{1+x}\right) ^{s}.  \label{Eq.6.6}
\end{equation}
Using the Euler transformation (\cite[p.33]{Srivastava3}): 
\begin{equation}
{}_{2}\digamma_{1}\left( {\QATOP{a,b}{c}}\big|z\right)
=(1-z)^{c-a-b}{}_{2}\digamma_{1}\left( {\QATOP{c-a,c-b}{c}}\big|z\right); \qquad
|\arg(1-z)|<\pi,  \label{EulerTransform}
\end{equation}
Eq.~\eqref{Eq.6.6} becomes 
\begin{equation}
\begin{split}
S_{n} = & \, (1+x)^{n}\left( \frac{x}{1+x}\right) ^{c-a-b} \\
& \times \sum_{s\geq 0}\frac{(c-a)_{s}(c-b)_{s}}{s!(c)_{s}}\left( \frac{x}{%
(1+x)^2}\right) ^{s}{}_{2}\digamma_{1}\left( {\QATOP{c-a+s,c-b+s}{c+s}}%
\bigg| \frac{1}{1+x}\right).
\end{split}
\label{A.08}
\end{equation}
Next, by applying the generating formula (\cite[p.348]{Prudnikov}): 
\begin{equation}  \label{A.09}
\sum_{k\geq 0}\frac{(\bar{a})_{k}(\bar{b})_{k}}{k!(c)_{k}}%
t^{k}{}_{2}\digamma_{1}\left( {\QATOP{\bar{a}+k,\bar{b}+k}{c+k}}\big|%
y\right) ={}_{2}\digamma_{1}\left( {\QATOP{\bar{a},\bar{b}}{c}}\big|%
t+y\right),\qquad |t|,\quad|y|,\quad|t+y|<1,
\end{equation}%
for $\bar{a}=c-a$,\quad$\bar{b}=c-b$,\quad$t=\frac{x}{(1+x)^2}$ and $y=\frac{1}{1+x}$%
, the sum $S_{n}$ can be written in a closed form as 
\begin{equation}
S_{n} = (1+x)^{n}\left( \frac{x}{1+x}\right)
^{c-a-b}{}_{2}\digamma_{1}\left( {\QATOP{c-a,c-b}{c}}\bigg|\frac{1+2x}{%
(1+x)^{2}}\right).  \label{A.10}
\end{equation}
where the conditions $|t|,\quad|y|,\quad|t+y|<1$ translate into $x\in]-\infty,-2\sqrt{2%
}[\cup]0,+\infty[$. Note that if $\max\{a,b\}\in\mathbb{Z}_{-}$ and $%
\max\{c-a,c-b\}\in\mathbb{Z}_{-}$, the convergence conditions $|t|,\quad|y|,
\quad|t+y|<1$ are not necessary for applying the formula in Eq.~\eqref{A.09} since
both series in L.H.S and R.H.S of Eq.~\eqref{A.08} will be finite sums. Then
the resulting Eq.~\eqref{A.10} holds true for all $x\in\mathbb{C}$ in this
case. This completes the proof. \hfill $\square $

\section{Concluding remarks}

\label{section7} We have been dealing with the Husimi Q-function of a
density operator with respect to a set of coherent states attached to higher
hyperbolic Landau levels and labeled by points of an open disk of radius $R$
(phase space). For both a pure state representing a projector on a Fock
state and the thermal density operator (mixed states) for the Hamiltonian of
the isotonic oscillator, we explicitly have established formulas for the
Q-function and we have written down the corresponding characteristic
function from which the mean statistical parameters have been derived. We
also have obtained a lower bound for a thermodynamical potential associated
with the Hamiltonian under consideration. Most of the results of the
Euclidean setting (flat case) as $R\to\infty$ have been recovered, by making
use of asymptotic formulas involving special functions and orthogonal
polynomials. In particular, and as tool, we obtained a summation formula for
the Kamp\'{e} de F\'eriet function $\digamma_{2:0:0}^{1:2:2}$. The obtained
results may be useful for tackling the hyperbolic version of Ginibre-type
processes \cite{Shirai}. They also can be exploited in the quantization
procedure or localization operators "\`{A} la Daubechies" \cite{Daubechies}%
. Finally, these Q-functions will be used to determine Wehrl entropies and
discuss some related questions in our forthcoming paper.

\subsection*{Acknowledgements}
The authors would like to thank the Moroccan Association of Harmonic Analysis \& Spectral Geometry.

\appendix

\section{Proof of Eq.~(4.5)}

\label{AppB} By the asymptotic expansion for the ratio of the Gamma
functions (\cite[p.24]{Srivastava3}): 
\begin{equation}
\frac{\Gamma (z+\alpha )}{\Gamma (z+\beta )}=z^{\alpha -\beta }\left( 1+ 
\frac{(\alpha -\beta )(\alpha +\beta -1)}{2z}+\mathcal{O}\left(
z^{-2}\right) \right)  \label{asympGamma}
\end{equation}
$|z|\rightarrow \infty ,\ |arg(z)|\leq \pi -\varepsilon ,\ 0<\varepsilon<\pi$%
, one obtains that 
\begin{equation}
\frac{\Gamma (2(BR^{2}-m)+m\vee j)}{\Gamma (2(BR^{2}-m)+m\wedge j)}\simeq
(2BR^{2})^{|m-j|}\text{ as\ }R\rightarrow \infty.  \label{B.1}
\end{equation}%
Next, using the well known approximation (\cite[p.196]{Rudin}): 
\begin{equation}
\left( 1+\frac{x}{n}\right) ^{n}\;\simeq \;e^{x}\text{ as }n\rightarrow
\infty  \label{ExpFormula}
\end{equation}%
we find that 
\begin{equation}
\left( 1-\frac{|z|^{2}}{R^{2}}\right) ^{2(BR^{2}-m)}= \left( 1-\frac{%
2B|z|^{2}}{2BR^{2}}\right) ^{2BR^{2}}\left( 1-\frac{|z|^{2}}{R^{2}}\right)
^{-2m} \simeq e^{-2B|z|^{2}}\text{ as }R\rightarrow \infty.  \label{B.2}
\end{equation}%
By another side, we use the limiting formula (\cite[p.247]{Magnus}): 
\begin{equation}
L_{n}^{(a )}(x)=\lim_{b \rightarrow \infty }P_{n}^{(a,b)} \left(1-\frac{2x}{b%
}\right)  \label{limitingLnPn}
\end{equation}
to show that 
\begin{equation}
P_{m\wedge j}^{(|m-j|,\,2(BR^{2}-m)-1)}\left( 1-\frac{2|z|^{2}}{R^{2}}%
\right) \simeq L_{m\wedge j}^{(|m-j|)}(2B|z|^{2})\text{ as }R\rightarrow
\infty.  \label{B.3}
\end{equation}%
Finally by Eq.~\eqref{B.1}, Eq.~\eqref{B.2} and Eq.~\eqref{B.3} we arrive at 
\begin{equation}
Q_{j}^{B,R,m}(z)\simeq \frac{(m\wedge j)!}{(m\vee j)!}\;e^{-2B|z|^{2}}\;%
\left( 2B|z|^{2}\right) ^{|m-j|}\left( L_{m\wedge
j}^{(|m-j|)}(2B|z|^{2})\right) ^{2}\text{ as }R\rightarrow \infty .
\label{B.5}
\end{equation}%
The expression in the R.H.S of Eq.~\eqref{B.5} is $Q_{j}^{\left( m\right) ,%
\text{ Euclid}}(\sqrt{2B}z)$ as denoted in Eq.~\eqref{3.6}.
\hfill $\square $

\section{Proof of Eq.~(4.14)}

\label{AppC} We start by recalling the characteristic function as defined in
Eq.~\eqref{CFPureDef} by 
\begin{equation}
\phi _{\mathfrak{X}_{m}^{\left( R\right) }}\left( u\right)
=\int\limits_{0}^{R^{2}}e^{iu\lambda }\mathfrak{Q}_{j}^{B,R,m} \left(
\lambda \right) d\lambda.
\end{equation}
By Eq.~\eqref{3.10} it can also be written as 
\begin{equation}
\phi _{\mathfrak{X}_{m}^{\left( R\right) }}\left( u\right) =
\left(2(BR^{2}-m)-1\right) \tau _{j}^{B,R,m} I_{R}  \label{C.2}
\end{equation}%
where $\tau _{j}^{B,R,m}$ is given in Eq.~\eqref{Tauj} and 
\begin{equation}
I_{R}:=\int_{0}^{1}e^{iuR^{2}x}\left( 1-x\right) ^{2(BR^{2}-m)-2}\left(
x\right) ^{|m-j|}\left( P_{m\wedge j}^{(|m-j|,2(BR^{2}-m)-1)}\left(
1-2x\right) \right) ^{2}dx.
\end{equation}%
Next, we introduce the notations $b=2(BR^{2}-m)-1$, $a=|m-j|$ and $l=m\wedge
j$ with the variable change $v=2x-1$ in order to rewrite $I_{R}$ as 
\begin{equation}
I_{R}=\frac{1}{2^{a+b}}\int_{-1}^{1}(v+1)^{a}(1-v)^{b-1}e^{\frac{iuR^{2}}{2}%
(v+1)}\left( P_{l}^{(a,b)}(-v)\right) ^{2}dv.  \label{C.4}
\end{equation}%
For the square of the Jacobi polynomial in the last integral we may use the
following Clebsh-Gordan type linearization formula (\cite[p.611]{Chaggara}): 
\begin{equation}
P_{i}^{(\lambda ,\delta )}(x)P_{j}^{(\mu ,\gamma
)}(x)=\sum_{k=0}^{i+j}C_{k}P_{k}^{(\alpha ,\beta )}(x)
\end{equation}%
where the coefficients 
\begin{equation}
\begin{split}
C_{k} = &\frac{(i+j)!(\alpha +1)_{i+j}(\lambda +\delta +1)_{2i}(\mu +\gamma
+1)_{2j}}{i!j!(\lambda +\delta +1)_{i}(\mu +\gamma +1)_{j}}\frac{%
(-1)^{i+j-k}(\alpha +\beta +1)_{k}(2k+\alpha +\beta +1)}{(i+j-k)!(\alpha
+1)_{k}(\alpha +\beta +1)_{i+j+k+1}} \\
& \times \digamma_{2:1:1}^{2:2:2}\left( 
\begin{array}{c}
-(i+j)+k,-\alpha -\beta -1-(i+j)-k;-i,-\lambda -i;-j,-\mu -j \\ 
-(i+j),-\alpha -(i+j);-2i-\lambda -\delta ;-2j-\mu -\gamma%
\end{array}
\big|1,1\right)
\end{split}%
\end{equation}
are chosen with parameters $\lambda =\mu =a$, $\delta =\gamma =b$ and $i=j=l$%
. Therefore, $I_R$ reads 
\begin{equation}
I_{R}=\frac{1}{2^{a+b}}\sum_{k=0}^{2l}C_{k}(-1)^{k}\int%
\limits_{-1}^{+1}(v+1)^{a}(1-v)^{b-1}e^{\frac{1}{2}iuR^{2}(v+1)}P_{k}^{(%
\beta ,\alpha )}(v)dv  \label{C.7}
\end{equation}%
with 
\begin{equation}
\begin{split}
C_{k} = &\frac{(2l)!(\alpha +1)_{2l}((a+b+1)_{2l})^{2}}{%
(l!)^{2}((a+b+1)_{l})^{2}}\frac{(-1)^{2l-k}(\alpha +\beta +1)_{k}(2k+\alpha
+\beta +1)}{(2l-k)!(\alpha +1)_{k}(\alpha +\beta +1)_{2l+k+1}} \\
& \times \digamma_{2:1:1}^{2:2:2}\left( 
\begin{array}{c}
-2l+k,-\alpha -\beta -1-2l-k;-l,-a-l;-l,-a-l \\ 
-2l,-\alpha -2l;-2l-a-b;-2l-a-b%
\end{array}%
\big|1,1\right).
\end{split}
\label{C.8}
\end{equation}%
Now, the integral in Eq.~\eqref{C.7} can be computed by applying the formula
(see \cite[p.572]{PrudnikovVol2}): 
\begin{equation*}
\int_{-\omega }^{\omega }(x+\omega )^{\gamma -1}(\omega -x)^{\rho
}e^{-p(x+\omega )^{r}}P_{n}^{(\rho ,\sigma )}(\frac{x}{\omega })dx=\frac{%
(-1)^{n}}{n!}(2\omega )^{\gamma +\rho }\Gamma (\rho +n+1)
\end{equation*}%
\begin{equation}
\times \sum_{s\geq 0}\frac{1}{s!}(\sigma -\gamma -rs+1)_{n}\frac{\Gamma
(\gamma +rs)}{\Gamma (\gamma +\rho +rs+n+1)}(-2^{r}\omega ^{r}p)^{s},
\label{C.9}
\end{equation}
$\omega ,r,\Re\gamma >0,\Re\rho >-1$ for parameters: $\omega =1,\gamma
=a+1=|m-j|+1,\rho =\beta =b-1=2(BR^{2}-m)-2,p=-iuR^{2}/2,r=1,\sigma =\alpha
=\gamma -1=|m-j|,n=k$. Summarizing the above calculations, Eq.~\eqref{C.2}
reads 
\begin{equation}
\begin{split}
\phi _{\mathfrak{X}_{m}^{\left( R\right) }}\left( u\right) =& \,
\left(2(BR^{2}-m)-1\right) \tau _{j}^{B,R,m} \sum_{k=0}^{2(m\wedge j)}
C_{j,k}^{B,R,m}\frac{(-1)^{k}}{k!} \Gamma (2(BR^{2}-m)-1+k) \\
& \times \sum_{s\geq 0}\frac{1}{s!}\frac{\Gamma (|m-j|+s+k+1)}{\Gamma
(2(BR^{2}-m)+|m-j|+s+2k)}(iuR^{2})^{s+k}
\end{split}
\label{C.10}
\end{equation}%
where 
\begin{equation}
\begin{split}
&C_{j,k}^{B,R,m} =\frac{(2(m\wedge j))!(|m-j|+1))_{2(m\wedge
j)}((2(BR^{2}-m)+|m-j|)_{2(m\wedge j)})^{2}}{((m\wedge
j)!)^{2}((2(BR^{2}-m)+|m-j|)_{m\wedge j})^{2}} \\
&\times \frac{(-1)^{k}(2(BR^{2}-m)+|m-j|-1)_{k}(2k+2(BR^{2}-m)+|m-j|-1)}{%
(2(m\wedge j)-k)!(|m-j|+1)_{k}(2(BR^{2}-m)+|m-j|-1)_{2(m\wedge j)+k+1}} \\
&\times \digamma_{2:1:1}^{2:2:2}\left( 
\begin{array}{c}
-2(m\wedge j)+k,-2(BR^{2}-m)-m-j+1-k;-m,-j;-m,-j \\ 
-2(m\wedge j),-m-j;-2(BR^{2}-m)-m-j+1;-2(BR^{2}-m)-m-j+1%
\end{array}%
\bigg|1,1\right)
\end{split}
\label{C.11}
\end{equation}
Finally, the series over $s$ in the R.H.S of Eq.~\eqref{C.10} can also be
written in terms of a $_{1}\digamma_{1}$-series. This ends the proof. \hfill 
$\square$

\section{Proof of Eq.~(4.19)}
\label{AppD} 
First, we note that by using the asymptotic
expansion Eq.~\eqref{asympGamma} 
\begin{equation}
\tau _{m,j}^{B,R}(2(BR^{2}-m)-1)\simeq \frac{(m\wedge j)!}{(m\vee j)!}%
(2BR^{2})^{|m-j|+1} \quad \text{as}\quad R\rightarrow \infty  \label{D.1}
\end{equation}%
and 
\begin{equation}
\frac{\Gamma (2(BR^{2}-m)-1+k)}{\Gamma (2(BR^{2}-m)+|m-j|+s+2k)}{\simeq }%
\frac{1}{(2BR^{2})^{1+|m-j|+s+k}} \quad \text{as}\quad R\rightarrow \infty .
\label{D.2}
\end{equation}
Next, we make us of the approximation formula which derive from Eq.~\eqref{asympGamma} 
\begin{equation}
(z)_{a }=z^{a }\left( 1+\frac{a (a -1)}{2z}+\mathcal{O}\left( z^{-2}\right)
\right), \;\; |z|\rightarrow \infty ,\ |arg(z)|\leq \pi -\varepsilon,\
0<\varepsilon<\pi  \label{asympPochammer}
\end{equation}%
to obtain 
\begin{eqnarray}
C_{j,k}^{B,R,m} &\simeq &\frac{(2(m\wedge j))!}{((m\wedge j)!)^{2}}\frac{
(|m-j|+1)_{2(m\wedge j)}(-1)^{k}}{(2(m\wedge j)-k)!(|m-j|+1)_{k}}  \label{C4}
\\
&&\times F_{2:1:1}^{2:2:2}\left( 
\begin{array}{c}
-2(m\wedge j)+k,-2BR^{2};-m,-j;-m,-j \\ 
-2(m\wedge j),-m-j;-2BR^{2};-2BR^{2}%
\end{array}%
\bigg|1,1\right) \quad \text{as}\quad R\rightarrow \infty .  \notag
\end{eqnarray}%
From the definition of $\digamma_{l:m:n}^{p:q:k}$ in Eq.~\eqref{kampe}, we
may rewrite Eq.~\eqref{C4} as

\begin{eqnarray}
C_{j,k}^{B,R,m} &\simeq &\frac{(2(m\wedge j))!}{((m\wedge j)!)^{2}}\frac{%
(|m-j|+1)_{2(m\wedge j)}(-1)^{k}}{(2(m\wedge j)-k)!(|m-j|+1)_{k}}
\label{D.4} \\
&&\times \digamma_{2:0:0}^{1:2:2}\left( {\QATOP{-2(m\wedge j)+k;-m,-j;-m,-j}{%
-2(m\wedge j),-m-j;-;-}}\bigg|1,1\right) \quad \text{as}\quad R\rightarrow
\infty.  \notag
\end{eqnarray}
Using equations Eq.~\eqref{D.1}, Eq.~\eqref{D.2} and Eq.~\eqref{D.4} we get,
after simplification, that 
\begin{eqnarray}
\phi _{\mathfrak{X}_{m}^{\left( R\right) }}\left( u\right) &\simeq &\frac{%
(m+j)!}{m!j!}\sum_{k=0}^{2(m\wedge j)}\binom{2(m\wedge j)}{k}%
\digamma_{2:0:0}^{1:2:2}\left( {\QATOP{-2(m\wedge j)+k;-m,-j;-m,-j}{%
-2(m\wedge j),-m-j;-;-}}\bigg|1,1\right)  \notag \\
&&\times \left( \frac{iu}{2B}\right) ^{k}\sum_{s\geq 0} (|m-j|+k+1)_{s} 
\frac{\left(\frac{iu}{2B}\right)^{s}}{s!} \quad \text{as}\quad R\rightarrow
\infty  \label{D.5}
\end{eqnarray}%
The infinite sum in Eq.~\eqref{D.5} reduces to 
\begin{equation}
\sum_{s\geq 0}\frac{1}{s!}(|m-j|+k+1)_{s}\left( \frac{iu}{2B}\right) ^{s}
=\left( 1-\frac{iu}{2B}\right) ^{-|m-j|-k-1}, \quad u<2B.
\end{equation}
Then for $u<2B$, Eq.~\eqref{D.5} becomes 
\begin{equation}
\begin{split}
\phi _{\mathfrak{X}_{m}^{\left( R\right) }}\left( u\right) & \simeq \frac{
(m+j)!}{m!j!}\left( 1-\frac{iu}{2B}\right) ^{-|m-j|-1} \\
& \times \sum_{k=0}^{2(m\wedge j)}\binom{2(m\wedge j)}{k}%
\digamma_{2:0:0}^{1:2:2}\left( {\QATOP{-2(m\wedge j)+k;-m,-j;-m,-j}{%
-2(m\wedge j),-m-j;-;-}}\bigg|1,1\right) \left( \frac{iu}{ 2B-iu}\right)
^{k},
\end{split}
\label{D.7}
\end{equation}
as $R\to \infty$. Applaying Proposition 6.1 for parameters $n=2(m\wedge j)$, 
$a=-m$, $b=-j$, $c=-m-j$ and $x=\frac{iu}{ 2B-iu}$, Eq.~\eqref{D.7} reduces
to 
\begin{equation}
\phi _{\mathfrak{X}_{m}^{\left( R\right) }}\left( u\right) {\simeq }\frac{
(m+j)!}{m!j!}\left( 1-\frac{iu}{2B}\right) ^{-(m+j+1)}{}_{2}\digamma_{1}
\left( {\QATOP{-m,-j}{-m-j}}\bigg|1+\left( \frac{u}{2B}\right) ^{2}\right),
\quad u<2B  \label{D.9}
\end{equation}
as $R\rightarrow \infty$ for $m\wedge j>0$. Then, we recover the announced limiting formula in Eq.~\eqref{CFPureConvergence} for $m\wedge j>0$ using the connexion formula in Eq.~\eqref{CFPureEuclidean0}. Now for $m\wedge j=0$, we get directly
\begin{equation}
\phi _{\mathfrak{X}_{m}^{\left(R\right)}}\left( u\right) = {}_1\digamma_1\left({m\vee j+1 \atop 2(BR^2-m)+m\vee j}\bigg|iuR^2\right) {\simeq } \left( 1-\frac{iu}{2B}\right) ^{-(m\vee j+1)},
\quad u<2B 
\end{equation}
as $R\to\infty$,
that coincides to Eq.~\eqref{CFPureConvergence} putting $m\wedge j=0$. This completes the proof. \hfill $\square$

\section{Proof of Eqs.~(4.16)-(4.17)}
\label{AppE} 
Let us write the characteristic function $\phi _{\mathfrak{X}%
_{m}^{\left( R\right) }}\left( u\right)$ as 
\begin{eqnarray}  \label{I1}
\phi _{\mathfrak{X}_{m}^{\left( R\right) }}\left( u\right) &=&
\sum_{k=0}^{2(m\wedge j)}\sum_{s\geq 0}\delta _{j,k,s}^{B,R,m}(iuR^{2})^{s+k}
\notag  \label{I0} \\
&=& \sum_{s\geq 0}\delta _{j,0,s}^{B,R,m}(iuR^{2})^{s}+\sum_{s\geq 0}\delta
_{j,1,s}^{B,R,m}(iuR^{2})^{s+1}+\sum_{k=2}^{2(m\wedge j)}\sum_{s\geq
0}\delta _{j,k,s}^{B,R,m}(iuR^{2})^{s+k}
\end{eqnarray}%
where 
\begin{equation}
\delta _{j,k,s}^{B,R,m}:=\tau _{m,j}^{B,R}(2(BR^{2}-m)-1)C_{j,k}^{B,R,m}%
\frac{(-1)^{k}}{k!s!}\frac{\Gamma(2(BR^{2}-m)-1+k) \Gamma (|m-j|+s+k+1)}{%
\Gamma (2(BR^{2}-m)+|m-j|+s+2k)}
\end{equation}%
$\tau _{m,j}^{B,R}$ and $C_{j,k}^{B,R,m}$ being given respectively by Eq.~%
\eqref{Tauj} and Eq.~\eqref{Cjk}. Then, the first derivative of $\phi _{%
\mathfrak{X}_{m}^{\left( R\right) }}\left( u\right)$ reads 
\begin{equation}
\begin{split}
\frac{d}{du} \phi _{\mathfrak{X}_{m}^{\left( R\right) }}\left( u\right) =&
\left( \delta _{j,0,1}^{B,R,m}+\delta _{j,1,0}^{B,R,m}\right) (iR^{2}) +
\sum_{s\geq 2}\delta _{j,0,s}^{B,R,m}s(iR^{2})(iuR^{2})^{s-1} \\
&+\sum_{s\geq 1}\delta _{j,1,s}^{B,R,m}(s+1)(iR^{2})(iuR^{2})^{s}
+\sum_{k=2}^{2(m\wedge j)} \sum_{s\geq
0}\delta_{j,k,s}^{B,R,m}(s+k)(iR^{2})(iuR^{2})^{s+k-1}.  \label{I.5}
\end{split}%
\end{equation}%
The mean value of $\mathfrak{X}_{m}^{(R)}$ is then given by 
\begin{equation}
\mathbb{E}\left(\mathfrak{X}_{m}^{\left( R\right) }\right) := -i\frac{d}{du}
\phi _{\mathfrak{X}_{m}^{\left( R\right) }} \bigg|_{u=0} = \left( \delta
_{j,0,1}^{B,R,m}+\delta _{j,1,0}^{B,R,m}\right) R^{2}.  \label{D.4Bis2}
\end{equation}
where 
\begin{equation}
\delta _{j,0,1}^{B,R,m} = \tau _{m,j}^{B,R}\frac{(|m-j|+1)!%
\Gamma(2(BR^{2}-m))} {\Gamma(2(BR^{2}-m)+|m-j|+1)}C_{j,0}^{B,R,m}
\end{equation}%
and 
\begin{equation}
\delta _{j,1,0}^{B,R,m} = -\tau _{m,j}^{B,R}\frac{(|m-j|+1)!%
\Gamma(2(BR^{2}-m))} {\Gamma(2(BR^{2}-m)+|m-j|+2)}%
(2(BR^{2}-m)-1)C_{j,1}^{B,R,m}.
\end{equation}
This completes the proof of Eq.~\eqref{EsperenceHusimiPure}. For $m=0$, it is
clear that $k=0$ and $C_{j,0}^{B,R,0}=1$. Then 
\begin{equation}
\mathbb{E}\left(\mathfrak{X}_{0}^{\left( R\right)}\right) =\kappa
_{j}^{B,R,0} \frac{(j+1)!\Gamma (2BR^{2})}{\Gamma (2BR^{2}+j+1)}R^{2} \\
= \frac{(j+1)R^{2}}{2BR^{2}+j}.
\end{equation}%
Now, for the variance, we search for the second derivative of $\phi _{%
\mathfrak{X} _{m}^{\left( R\right) }}\left( u\right)$. We rewrite Eq.~%
\eqref{I.5} as 
\begin{equation}
\begin{split}
\frac{d}{du} \phi _{\mathfrak{X}_{m}^{\left( R\right) }}\left( u\right)
=&\left( \delta _{j,0,1}^{B,R,m}+\delta _{j,1,0}^{B,R,m}\right)
(iR^{2})+\left( \delta _{j,0,2}^{B,R,m}+\delta _{j,1,1}^{B,R,m}+\delta
_{j,2,0}^{B,R,m}\right) 2(iR^{2})(iuR^{2}) \\
&+\sum_{s\geq 3}\delta _{j,0,s}^{B,R,m}s(iR^{2})(iuR^{2})^{s-1} +
\sum_{s\geq 2}\delta _{j,1,s}^{B,R,m}(s+1)(iR^{2})(iuR^{2})^{s} \\
&+\sum_{s\geq 1}\delta _{j,2,s}^{B,R,m}(s+2)(iR^{2})(iuR^{2})^{s+1} +
\sum_{k=3}^{2(m\wedge j)}\sum_{s\geq 0}\delta
_{j,k,s}^{B,R,m}(s+k)(iR^{2})(iuR^{2})^{s+k-1}.
\end{split}%
\end{equation}
Then, 
\begin{equation}
\begin{split}
\frac{d^{2}}{du^{2}} \phi _{\mathfrak{X}_{m}^{\left( R\right) }}(u) =&\left(
\delta _{j,0,2}^{B,R,m}+\delta _{j,1,1}^{B,R,m}+\delta
_{j,2,0}^{B,R,m}\right) 2(iR^{2})^{2} + \sum_{s\geq 3}\delta_{j,0,s}
^{B,R,m}(s-1)s(iR^{2})^{2}(iuR^{2})^{s-2} \\
&+\sum_{s\geq 2}\delta _{j,1,s}^{B,R,m}s(s+1)(iR^{2})^{2}(iuR^{2})^{s-1} +
\sum_{s\geq 1}\delta _{j,2,s}^{B,R,m}(s+1)(s+2)(iR^{2})^{2}(iuR^{2})^{s} \\
&+\sum_{k=3}^{2(m\wedge j)}\sum_{s\geq 0}\delta
_{j,k,s}^{B,R,m}(s+k-1)(s+k)(iR^{2})^{2}(iuR^{2})^{s+k-2}.
\end{split}%
\end{equation}
The evaluation at $u=0$ with the mean value gives the variance of $\mathfrak{%
X}_{m} ^{\left(R\right)}$ as 
\begin{equation}
\mathbb{V}ar\left( \mathfrak{X}_{m}^{\left( R\right) } \right) := -\frac{%
d^{2}} {du^{2}} \phi _{\mathfrak{X}_{m}^{\left( R\right) }}\left( u\right) %
\bigg|_{u=0}- \mathbb{E}\left( \mathfrak{X}_{m}^{\left( R\right) }
\right)^{2} = \left(\delta _{j,0,2}^{B,R,m}+\delta _{j,1,1}^{B,R,m}+\delta
_{j,2,0}^{B,R,m}\right) 2R^{4}- \mathbb{E}\left( \mathfrak{X}_{m}^{\left(
R\right) } \right)^{2}.  \label{D.10}
\end{equation}
Replacing 
\begin{equation}
\delta _{j,0,2}^{B,R,m} =\tau _{m,j}^{B,R}\frac{(|m-j|+2)!\Gamma(2(BR^{2}-m))%
} {2\Gamma(2(BR^{2}-m)+|m-j|+2)}C_{j,0}^{B,R,m},
\end{equation}
\begin{equation}
\delta_{j,1,1}^{B,R,m} = - \tau_{m,j}^{B,R} \frac{(|m-j|+2)!\Gamma(2(BR^2-m))%
} {\Gamma(2(BR^2-m)+|m-j|+3)} (2(BR^2-m)-1) C_{j,1}^{B,R,m}
\end{equation}
and 
\begin{equation}
\delta_{j,2,0}^{B,R,m} = \tau_{m,j}^{B,R} \frac{(|m-j|+2)!\Gamma(2(BR^2-m)+1)%
} {2\Gamma(2(BR^2-m)+|m-j|+4)} (2(BR^2-m)-1) C_{j,2}^{B,R,m}
\end{equation}
by theirs expressions into Eq.~\eqref{D.10} gives the announced formula in Eq.~%
\eqref{VarHusimiPure}. Like above, we find for $m=0$ 
\begin{eqnarray}
\mathbb{V}ar\left(\mathfrak{X}_{0}^{(R)}\right) &=& \kappa_{j}^{B,R,0} \frac{%
(j+2)!\Gamma(2BR^2)}{\Gamma(2BR^2+j+2)} \left(C_{j,0}^{B,R,0} \right)R^4 -
\left(\kappa_{j}^{B,R,0} \frac{(j+1)!\Gamma(2BR^2)}{ \Gamma(2BR^2+j+1)}
R^2\right)^2,  \notag \\
&=& \frac{(j+1)(2BR^2-1)R^4}{ (2BR^2+j)^2(2BR^2+j+1)}.
\end{eqnarray}
This ends the proof. \hfill $\square$

\section{Proof of Eq.~(5.24)}

\label{AppF} With the approximation Eq.~\eqref{ExpFormula} we find 
\begin{equation}
\left(\frac{1-\frac{z\bar{z}}{R^2}}{1-\frac{z\bar{z}}{R^2}e^{-\beta}}%
\right)^{2BR^2} = \left(1-\frac{2Bz\bar{z}}{2BR^2} \right)^{2BR^2} \left(1+%
\frac{2Bz\bar{z}e^{-\beta}}{-2BR^2}\right)^{-2BR^2} \simeq \exp(-2Bz\bar{z}%
(1-e^{-\beta}))  \label{E.1}
\end{equation}
as $R\to\infty$. Direct calculation gives 
\begin{equation}
\left(\frac{\left(\frac{z\bar{z}}{R^2}-e^{-\beta}\right)\left(1-\frac{z\bar{z%
}} {R^2}e^{-\beta}\right)}{\left(1-\frac{z\bar{z}}{R^2}\right)^2}\right)^m
\eqsim (-1)^m e^{-\beta m} \quad \text{as}\quad R \to \infty,  \label{E.2}
\end{equation}
and 
\begin{equation}
P^{(2(BR^2-m)-1,0)}_m \left(1+\frac{2e^{-\beta}\left(1-\frac{z\bar{z}}{R^2}
\right)^2}{ \left(\frac{z\bar{z}}{R^2}-e^{-\beta}\right) \left(1-\frac{z\bar{%
z}}{R^2}e^{-\beta}\right)}\right) \simeq (-1)^m P^{(0,2BR^2)}_m
\left(1\right), \quad \text{as}\quad R \to \infty
\end{equation}
that becomes 
\begin{equation}
P^{(2(BR^2-m)-1,0)}_m \left(1+\frac{2e^{-\beta}\left(1-\frac{z\bar{z}}{R^2}%
\right)^2}{ \left(\frac{z\bar{z}}{R^2}-e^{-\beta}\right)\left(1-\frac{z\bar{z%
}}{R^2}e^{-\beta}\right)}\right) \simeq (-1)^m L_m^{(0)}\left(-8Bz\bar{z}%
\sinh^2\frac{ \beta}{2}\right), \quad \text{as}\quad \to \infty  \label{E.4}
\end{equation}
by using the limiting formula Eq.~\eqref{limitingLnPn}. Then, it follow from
Eq.~\eqref{E.1}, Eq.~\eqref{E.2} and Eq.~\eqref{E.4} that 
\begin{equation}
Q_{\beta}^{B,R,m}(z) \simeq \left(1-e^{-\beta}\right) \exp\left(-2Bz\bar{z}
(1-e^{-\beta})\right) e^{-\beta m} L_m^{(0)}\left(-8Bz\bar{z}\sinh^2\frac{%
\beta}{2}\right), \quad \text{as}\quad R \to \infty.  \label{E.5}
\end{equation}
Comparing Eq.~\eqref{E.5} to \cite[Eq.~(3.14)]{Mouayn7}, we recover the limiting formula $%
Q_{\beta}^{B,R,m}(z) \to q_{m}(|\sqrt{2B}z|^2,\beta)$ as $R\to \infty$. \hfill $\square$

\section{Proof of Eq.~(5.28)}

\label{CFQMixedP} Identify $\alpha=2(BR^2-m)-1$ and $\xi=e^{-\beta}$ and set 
$x=\frac{\lambda}{R^2}$. Then, the characteristic function reads from Eq.~%
\eqref{4.22} as 
\begin{equation}
\begin{split}
\phi_{\mathfrak{Y}_{m}^{\left(R\right)}}( u) = & \alpha(1-\xi)
\int\limits_{0}^{1} e^{iuxR^2} \left(\frac{1-x}{1-x\xi}\right)^{\alpha+2m+1}
\left(\frac{(x-\xi)(1-x\xi)}{(1-x)^2}\right)^{m} \\
& \times P^{(\alpha,0)}_m \left(1+\frac{2\xi(1-x)^2}{(x-\xi)(1-x\xi)}\right) 
\frac{dx}{(1-x)^{2}}.
\end{split}
\label{CFQMixedP1}
\end{equation}
By using the Jacobi polynomial definition in Eq.~\eqref{jacobi} one obtains
the contraction 
\begin{eqnarray}
\left(\frac{(x-\xi)(1-x\xi)}{(1-x)^2}\right)^{m} P^{(\alpha,0)}_m \left(1+%
\frac{2\xi(1-x)^2}{(x-\xi)(1-x\xi)}\right) & = & \xi^m \sum_{k=0}^{m} \binom{%
m+\alpha}{k} \binom{m}{k} \left(\frac{(1-\xi)^2}{\xi}\right)^k  \notag \\
& &\times \left(\frac{x}{(1-x)^2}\right)^k.
\end{eqnarray}
Then, Eq.~\eqref{CFQMixedP1} becomes after reorganization 
\begin{equation}
\begin{split}
\phi_{\mathfrak{Y}_{m}^{\left(R\right)}}( u) = & \frac{\alpha(1-\xi)\xi^{m}}{%
(-\xi)^{\alpha+2m+1}} \\
& \sum_{k=0}^{m} \binom{m+\alpha}{k} \binom{m}{k} \left(\frac{(1-\xi)^2}{\xi}%
\right)^k \int\limits_{0}^{1} e^{iuxR^2} x^k (1-x)^{\alpha+2m-2k-1}
(x-\xi^{-1})^{-\alpha-2m-1} dx.
\end{split}
\label{CFQMixedP2}
\end{equation}
Thanks to the formula (\cite[p.328]{PrudnikovVol1}): 
\begin{equation}
\int\limits_{0}^{a} x^{\tau-1} (a-x)^{\beta-1} (x+z)^{-\rho} e^{-px} dx = 
\mathcal{B}(\tau, \beta) z^{-\rho} a^{\tau+\beta-1} \Phi_1\left( {\QATOP{%
\tau,\rho}{\tau+\beta}}\bigg|-\frac{a}{z}, ap \right)
\end{equation}
where $a, \Re \tau, \Re \beta > 0; |\arg(1+a/z)|<\pi$. It application on the
integral at the R.H.S of Eq.~\eqref{CFQMixedP2} for $a=1$, $z=-\xi^{-1}$, $%
\tau=k+1$, $\beta=\alpha+2m-2k$, $\rho=\alpha+2m+1$ and $p=-iuR^2$ gives
after simplification 
\begin{equation}
\begin{split}
\phi_{\mathfrak{Y}_{m}^{\left(R\right)}}( u) = & \, \alpha(1-\xi)\xi^{m}
\sum_{k=0}^{m} \binom{m+\alpha}{k} \binom{m}{k} \left(\frac{(1-\xi)^2}{\xi}%
\right)^k \\
& \mathcal{B}(k+1, \alpha+2m-2k) \Phi_1\left({\QATOP{k+1, \alpha+2m+1 }{%
\alpha+2m-k+1}} \Big| \xi, -iuR^2\right).
\end{split}
\label{CFQMixedP3}
\end{equation}
Replacing $\alpha$ and $\xi$ by their expressions ends the proof. \hfill $%
\square$

\section{Proof of Eqs.~(5.29)-(5.30)}

\label{EVQMixedP} Due to the definition of the Humbert series in Eq.~\eqref{Phi1Def}, it derivative reads 
\begin{equation}
\frac{\partial}{\partial z} \Phi_1\left({\QATOP{a,b }{c}} \Big| w, z\right)
= \frac{a}{c} \Phi_1\left({\QATOP{a+1,b }{c+1}} \Big| w, z\right).
\label{DPhi1}
\end{equation}
Then, the derivative of $\phi_{\mathfrak{Y}_{m}^{\left(R\right)}}( u)$ in Eq.~%
\eqref{CFQMixedP3} is given by 
\begin{eqnarray}
\frac{d}{du} \phi_{\mathfrak{Y}_{m}^{\left(R\right)}}( u) &=& \alpha
(1-\xi)\xi^{m} (-iR^{2}) \sum_{k=0}^{m} \binom{m+\alpha}{k} \binom{m}{k}
\left(\frac{(1-\xi)^2}{\xi}\right)^k  \notag \\
& & \mathcal{B}(k+2,\alpha+2m-2k) \Phi_1\left({\QATOP{k+2, \alpha+2m+1 }{%
\alpha+2m-k+2}} \Big| \xi, -iuR^2\right).  \label{DCFQMixedP}
\end{eqnarray}
The evaluation at $u=0$ taking into account the formula Eq.~\eqref{Phiw0}
gives the mean value of $\mathfrak{Y}_{m}^{\left(R\right)}$ as 
\begin{eqnarray}
\mathbb{E}\left(\mathfrak{Y}_{m}^{\left(R\right)}\right) &=& -i \frac{d}{du}
\phi_{\mathfrak{Y}_{m}^{\left(R\right)}}( u) \bigg|_{u=0}  \notag \\
&=& \alpha (\xi-1)\xi^{m} R^{2} \sum_{k=0}^{m} \binom{m+\alpha}{k} \binom{m}{%
k} \left(\frac{(1-\xi)^2}{\xi}\right)^k  \notag \\
& & \mathcal{B}(k+2,\alpha+2m-2k) {}_2\digamma_1\left({\QATOP{k+2,
\alpha+2m+1 }{\alpha+2m-k+2}} \Big| \xi\right).
\end{eqnarray}
For the variance, we search for the second derivative of $\phi_{\mathfrak{Y}%
_{m}^{(R)}}(u)$ that reads from Eq.~\eqref{DPhi1} and Eq.~\eqref{DCFQMixedP}
as 
\begin{eqnarray}
\frac{d^2}{du^2} \phi_{\mathfrak{Y}_{m}^{\left(R\right)}}( u) &=& -\alpha
(1-\xi)\xi^{m} R^{4} \sum_{k=0}^{m} \binom{m+\alpha}{k} \binom{m}{k} \left(%
\frac{(1-\xi)^2}{\xi}\right)^k  \notag \\
& & \mathcal{B}(k+3,\alpha+2m-2k) \Phi_1\left({\QATOP{k+3, \alpha+2m+1 }{%
\alpha+2m-k+3}} \Big| \xi, -iuR^2\right).
\end{eqnarray}
The evaluation at $u=0$ taking into account the formula Eq.~\eqref{Phiw0}
gives 
\begin{eqnarray}
\frac{d^2}{du^2} \phi_{\mathfrak{Y}_{m}^{\left(R\right)}}( u) \bigg|_{u=0}
&=& -\alpha (1-\xi)\xi^{m} R^{4} \sum_{k=0}^{m} \binom{m+\alpha}{k} \binom{m%
}{k} \left(\frac{(1-\xi)^2}{\xi}\right)^k  \notag \\
& & \mathcal{B}(k+3,\alpha+2m-2k) {}_2\digamma_1\left({\QATOP{k+3,
\alpha+2m+1 }{\alpha+2m-k+3}} \Big| \xi\right).
\end{eqnarray}
The variance of $\mathfrak{Y}_{m}^{\left(R\right)}$ that is defined by 
\begin{equation}
\mathbb{V}ar\left(\mathfrak{Y}_{m}^{\left(R\right)}\right) := - \frac{d^2}{%
du^2} \phi_{\mathfrak{Y}_{m}^{\left(R\right)}}( u) - \left(\mathbb{E}\left(%
\mathfrak{Y}_{m}^{\left(R\right)}\right)\right)^2
\end{equation}
reads then as 
\begin{equation*}
\mathbb{V}ar\left(\mathfrak{Y}_{m}^{\left(R\right)}\right) =
\end{equation*}
\begin{equation}
\begin{split}
&\, \alpha (1-\xi)\xi^{m} R^{4} \sum_{k=0}^{m} \binom{m+\alpha}{k} \binom{m}{%
k} \left(\frac{(1-\xi)^2}{\xi}\right)^k \mathcal{B}(k+3,\alpha+2m-2k)
{}_2\digamma_1\left({\QATOP{k+3, \alpha+2m+1 }{\alpha+2m-k+3}} \Big| %
\xi\right) - \\
& \Bigg[\alpha (1-\xi)\xi^{m} R^{2} \sum_{k=0}^{m} \binom{m+\alpha}{k} 
\binom{m}{k} \left(\frac{(1-\xi)^2}{\xi}\right)^k \mathcal{B}%
(k+2,\alpha+2m-2k) {}_2\digamma_1\left({\QATOP{k+2, \alpha+2m+1 }{%
\alpha+2m-k+2}} \Big| \xi\right) \Bigg]^2.
\end{split}%
\end{equation}
This ends the proof replacing $\alpha$ and $\xi$ by their expressions.
\hfill $\square$

\section{Proof of Eq.~(5.31)}

\label{LimitCFQMixedP} By using the formula Eq.~\eqref{asympGamma}, we obtain 
\begin{equation}
\binom{2BR^{2}-m-1}{k}\mathcal{B}(k+1,2BR^{2}-2k-1)=\frac{\Gamma (2BR^{2}-m)%
}{\Gamma (2BR^{2}-m-k)}\frac{\Gamma (2BR^{2}-2k-1)}{\Gamma (2BR^{2}-k)}%
\simeq \frac{1}{2BR^{2}}
\end{equation}%
as $R\rightarrow +\infty $, that yields 
\begin{equation}
(2(BR^{2}-m)-1)\binom{2BR^{2}-m-1}{k}\mathcal{B}(k+1,2BR^{2}-2k-1)\simeq
1, \quad \text{as}\quad R\rightarrow \infty .  \label{D3}
\end{equation}%
Next, by the definition of $\Phi _{1}$ in Eq.~\eqref{Phi1Def} and the use of
the approximation formula Eq.~\eqref{asympPochammer} we obtain after
simplification 
\begin{equation}
\Phi _{1}\left( {\QATOP{k+1,2BR^{2}}{2BR^{2}-k}}\Big|e^{-\beta
},-iuR^{2}\right) \simeq \sum_{s,t=0}^{+\infty }(k+1)_{s+t}\frac{e^{-s\beta }%
}{s!}\frac{(\frac{iu}{2B})^{t}}{t!}, \quad \text{as}\quad R\rightarrow \infty .
\label{D4}
\end{equation}%
By the use of the fact $(a)_{n+m}=(a)_{n}(a+n)_{m}$, the series at the R.H.S
of Eq.~\eqref{D4} becomes 
\begin{equation}
\sum_{s,t=0}^{+\infty }(k+1)_{s+t}\frac{e^{-s\beta }}{s!}\frac{(\frac{iu}{2B}%
)^{t}}{t!}=\sum_{s=0}^{+\infty }(k+1)_{s}\frac{e^{-s\beta }}{s!}%
\sum_{t=0}^{+\infty }(k+1+s)_{t}\frac{(\frac{iu}{2B})^{t}}{t!}.
\label{D.4Bis}
\end{equation}%
Identifying the series (over t) at the R.H.S of Eq.~\eqref{D.4Bis} as a $%
{}_{1}\digamma _{0}$, it follow for $u<2B$ 
\begin{eqnarray}
\sum_{s,t=0}^{+\infty }(k+1)_{s+t}\frac{e^{-s\beta }}{s!}\frac{(\frac{iu}{2B}%
)^{t}}{t!} &=&\sum_{s=0}^{+\infty }(k+1)_{s}\frac{e^{-s\beta }}{s!}\left( 1-%
\frac{iu}{2B}\right) ^{-k-1-s}  \notag \\
&=&\left( 1-\frac{iu}{2B}\right) ^{-k-1}\sum_{s=0}^{+\infty }(k+1)_{s}\frac{%
[(1-\frac{iu}{2B})^{-1}e^{-\beta }]^{s}}{s!}
\end{eqnarray}%
that becomes identifying twice ${}_{1}\digamma _{0}$ as 
\begin{eqnarray}
\sum_{s,t=0}^{+\infty }(k+1)_{s+t}\frac{e^{-s\beta }}{s!}\frac{(\frac{iu}{2B}%
)^{t}}{t!} &=&\left( 1-\frac{iu}{2B}\right) ^{-k-1}\left( 1-(1-\frac{iu}{2B}%
)^{-1}e^{-\beta }\right) ^{-k-1}  \notag \\
&=&\left( 1-e^{-\beta }-\frac{iu}{2B}\right) ^{-k-1}.
\end{eqnarray}%
Then, Eq.~\eqref{D4} becomes 
\begin{equation}
\Phi _{1}\left( {\QATOP{k+1,2BR^{2}}{2BR^{2}-k}}\Big|e^{-\beta
},-iuR^{2}\right) \simeq \left( 1-e^{-\beta }-\frac{iu}{2B}\right)
^{-k-1},\quad \text{as}\quad R\rightarrow \infty   \label{D8}
\end{equation}%
for $u<2B$. Thus, from Eq.~\eqref{D3} and Eq.~\eqref{D8}, $\phi _{\mathfrak{Y}%
_{m}^{\left( R\right) }}(u)$ has as limit when $R\rightarrow \infty $ 
\begin{equation}
\phi _{\mathfrak{Y}_{m}^{\left( R\right) }}(u)\simeq (1-e^{-\beta
})e^{-m\beta }\sum_{k=0}^{m}\binom{m}{k}\left( \frac{(1-e^{-\beta })^{2}}{%
e^{-\beta }}\right) ^{k}\left( 1-e^{-\beta }-\frac{iu}{2B}\right)
^{-k-1},\quad u<2B
\end{equation}%
that reads successively 
\begin{eqnarray}
\phi _{\mathfrak{Y}_{m}^{\left( R\right) }}(u) &\simeq &\frac{(1-e^{-\beta
})e^{-m\beta }}{1-e^{-\beta }-\frac{iu}{2B}}\sum_{k=0}^{m}\binom{m}{k}\left( 
\frac{\frac{(1-e^{-\beta })^{2}}{e^{-\beta }}}{1-e^{-\beta }-\frac{iu}{2B}}%
\right) ^{k}  \notag \\
&\simeq &\frac{(1-e^{-\beta })e^{-m\beta }}{1-e^{-\beta }-\frac{iu}{2B}}%
\left( 1+\frac{\frac{(1-e^{-\beta })^{2}}{e^{-\beta }}}{1-e^{-\beta }-\frac{%
iu}{2B}}\right) ^{m}  \notag \\
&\simeq &\frac{(1-e^{-\beta })}{1-e^{-\beta }-\frac{iu}{2B}}\left( \frac{%
1-e^{-\beta }-\frac{iu}{2B}e^{-\beta }}{1-e^{-\beta }-\frac{iu}{2B}}\right)
^{m},\quad u<2B.  \label{H.09}
\end{eqnarray}%
This ends the proof by comparing Eq.~\eqref{H.09} with (\cite[Eq.~(3.24)]{Mouayn7}). \hfill $\square$

\end{document}